\documentclass{article}

\usepackage{graphicx}
\usepackage{enumerate}
\usepackage{natbib}
\usepackage{url} %

\usepackage{amsmath}
\usepackage{amssymb}
\usepackage{amsthm}
\usepackage{hyperref}
\usepackage{algorithm,algpseudocode}
\usepackage{dsfont}
\usepackage[dvipsnames]{xcolor} 
\usepackage{booktabs}
\usepackage[affil-it]{authblk}
\usepackage{cleveref}
\usepackage{tabularx}
\usepackage{authblk}
\usepackage{enumitem}
\usepackage{stmaryrd}
\usepackage{bm}
\usepackage{orcidlink}

\usepackage{todonotes}
\usepackage{subfigure}
\usepackage{adjustbox}
\usepackage{multirow}
\usepackage{amsmath}
\usepackage{mathtools}
\usepackage{setspace}

\newcommand{\bs}{\boldsymbol}

\newcommand{\E}{\operatorname{E}}

\newcommand{\T}{{\!\mathsf{T}}}  %

\newcommand{\x}{\mathbf{x}}

\newcommand{\y}{\mathbf{y}}
  
\newcommand{\K}{\mathbf{K}}
\newcommand{\M}{\mathbf{M}}
\newcommand{\C}{\mathbf{C}}

\newcommand{\Q}{\mathbf{Q}}

\newcommand{\U}{\mathbf{U}}

\newcommand{\X}{\mathbf{X}} 
\newcommand{\Y}{\mathbf{Y}}

\theoremstyle{plain}

\newtheorem{theorem}{Theorem}[section]

\theoremstyle{definition}
\newtheorem{definition}[theorem]{Definition}

\title{Stochastic Variational Inference for Vine Copula Distributional Regression}
\author[1]{Gianmarco Callegher\,\orcidlink{0009-0001-1020-7887}}
\author[1]{Thomas Kneib\,\orcidlink{0000-0003-3390-0972}}

\affil[1]{University of G\"ottingen}
\date{\today}

\begin{document}

\maketitle

\begin{abstract}
Structured additive distributional regression flexibly relates all parameters of a conditional response distribution to covariates, but multivariate extensions remain challenging when dependence is complex. We propose a multivariate structured additive distributional regression model based on regular vine copulas. Different vine edges may use different pair-copula families, while every pair-copula parameter may vary with covariates through a structured additive predictor. The model therefore accommodates heterogeneous marginal distributions together with pair-specific asymmetric, tail-dependent, and covariate-dependent dependence structures.

For scalable inference, we develop a tree-wise stochastic variational inference procedure based on component-specific Gaussian variational approximations. Marginal models are estimated first, followed by pair-copula regressions sequentially along the vine trees. We also adapt sequential vine selection by fitting candidate covariate-dependent pair-copula regressions and using information criteria based on effective degrees of freedom for both family and tree selection. In simulations, the tree-wise estimator remains close to an oracle using the true recursive conditional inputs, whereas global refinement yields more concentrated approximations and lower frequentist coverage for upstream components. An application to six-dimensional meteorological data from the Netherlands yields a vine combining Gaussian and non-Gaussian pair copulas, with pronounced nonlinear spatial and temporal variation in dependence.
\end{abstract}

\section{Introduction}\label{sec:introduction}

Multivariate responses arise naturally whenever several outcomes describe different aspects of the same observational unit. Meteorological variables are measured jointly at the same location and time, pollutants co-occur under common atmospheric conditions, and economic indicators evolve together in response to shared external drivers. In such settings, modelling each response separately is generally insufficient. Even when the marginal distributions are estimated accurately, separate regressions do not determine the probability of joint events or the way in which the association between the responses changes across the covariate space. A coherent multivariate regression model must therefore describe both the conditional distribution of each response component and their dependence after conditioning on the observed covariates.

Structured additive distributional regression (SADR) provides a flexible framework for modelling conditional response distributions. Rather than restricting covariate effects to the conditional mean, each parameter of a chosen response distribution is related to a structured additive predictor \citep{rigby2005generalized,kneib2021}. Covariates may therefore affect several features of the conditional distribution simultaneously. Within a Bayesian treatment, suitable priors on regression coefficients and smoothing parameters provide regularization, while posterior inference enables uncertainty quantification for distributional parameters and covariate effects. Most methodological developments in SADR have focused on univariate responses, while flexible copula-based extensions have often been restricted to the bivariate case.

An early general extension to multivariate responses was proposed by \citet{klein2015bayesian}, who developed structured additive distributional regression for multivariate parametric response distributions and allowed each distributional parameter to depend on covariates. Within copula-based distributional regression, much of the subsequent methodological development has focused on bivariate responses. \citet{klein2016simultaneous} proposed a Bayesian framework for structured additive conditional copula regression, \citet{marra2017bivariate} developed bivariate copula additive models based on penalized likelihood inference, and \citet{hans2023boosting} introduced model-based boosting for bivariate distributional copula regression. Going beyond a single bivariate copula, \citet{vatter2018generalized} developed generalized additive models for pair-copula constructions, allowing the parameters of individual pair copulas in a vine to vary with covariates.

Several related contributions extend multivariate distributional regression beyond the bivariate case. Multivariate conditional transformation models describe flexible conditional marginal distributions together with a covariate-dependent Gaussian copula \citep{klein2022multivariate}. Cholesky-based multivariate Gaussian regression relates the means and the entries of a Cholesky representation of the covariance matrix to additive predictors, thereby extending Gaussian distributional regression to higher-dimensional response vectors \citep{muschinski2024cholesky}. More recently, \citet{kock2023truly} proposed a truly multivariate structured additive distributional regression model that combines potentially different marginal distributions through a Gaussian copula with a covariate-dependent correlation matrix.

These approaches provide coherent models for response dimensions beyond the bivariate case, but their dependence structures remain Gaussian. Cholesky-based multivariate Gaussian regression directly assumes a multivariate Gaussian response distribution, whereas multivariate conditional transformation models and the model of \citet{kock2023truly} describe dependence through a Gaussian copula. After transforming the conditional marginals to latent Gaussian variables, dependence is characterized by correlations on the latent Gaussian scale. These correlations may vary nonlinearly with the covariates, but the conditional copula remains Gaussian at every covariate value. Consequently, upper- and lower-tail dependence cannot be modelled asymmetrically and, apart from degenerate cases, the Gaussian copula is tail independent. It therefore cannot describe, for example, lower-tail dependence for one pair of responses, upper-tail dependence for another pair, and tail-independent association for a third pair within the same multivariate model.

Vine copulas provide a substantially more flexible alternative. A vine decomposes a multivariate copula density into a product of bivariate unconditional and conditional copula densities \citep{bedford2001probability,bedford2002vines,aas2009,joe2014dependence,czado2022vine}. Each pair-copula component may be selected from a different parametric family and can therefore represent a different form of association. By organizing these pair copulas in a sequence of linked trees, regular vines construct flexible multivariate distributions from bivariate building blocks. In a distributional regression setting, both the parameters of the conditional marginal distributions and those of the pair copulas can depend on covariates through structured additive predictors.

The resulting model is truly multivariate in two complementary senses. First, it applies to response vectors of arbitrary dimension and permits different marginal distributions for the individual response components. Second, dependence is flexible both across and within vine edges. Different edges may employ different pair-copula families and can therefore represent distinct forms of asymmetric, tail-dependent, or tail-independent association. At the same time, every parameter of every pair copula may depend on covariates through a structured additive predictor, allowing the strength and shape of these pair-specific dependence structures to vary over the covariate space.

This flexibility creates a substantial inferential challenge. A complete regular vine in dimension $D$ contains $D(D-1)/2$ pair copulas in addition to the $D$ marginal models, and each component may contain several parameters and multiple structured additive effects. Moreover, higher-tree likelihood contributions depend recursively on conditional probability integral transforms obtained from preceding components. Bayesian inference for such models can therefore become computationally demanding, particularly for large samples and during vine selection, where several candidate pair-copula regressions may have to be fitted for every admissible edge.

We consider stochastic variational inference as a scalable alternative. Variational inference approximates the posterior distribution within a tractable parametric family and determines the parameters of this approximation through optimization \citep{blei2017variational}, while stochastic variational inference permits likelihood contributions and their gradients to be estimated from mini-batches of observations \citep{hoffman2013stochastic}. Building on recent work on stochastic variational inference for structured additive distributional regression \citep{callegher2024stochastic}, we use, for each marginal and pair-copula regression, a multivariate Gaussian variational approximation for the regression coefficients and independent univariate Gaussian variational approximations for the log-smoothing variances.

We exploit the recursive structure of a regular vine through tree-wise estimation. Marginal distributional regressions are estimated first, and their fitted conditional distribution functions transform the observations to the copula scale. Pair-copula regressions are then estimated sequentially along the vine trees. After estimating a tree, its fitted $h$-functions, defined as partial derivatives of the corresponding pair-copula distribution functions and representing conditional distribution functions, provide the conditional pseudo-observations required by the next tree.

A related tree-wise idea has recently been proposed by \citet{griesbauer2026stepwise} in a different variational inference setting. They construct a variational posterior whose univariate margins are coupled through a D-vine copula and estimate the parameters governing this posterior dependence sequentially along the vine trees. In our setting, by contrast, the vine is part of the response likelihood, while the posterior distributions of the marginal and pair-copula regression parameters are approximated by Gaussian variational distributions. We use the standard ELBO rather than the R\'enyi objective considered by \citet{griesbauer2026stepwise}. Their setting concerns a vine copula used within the variational posterior, whereas the vine in the present model enters the response likelihood. Moreover, although the rescaled observation-level mini-batch log-likelihood is unbiased, substituting it into the nonlinear R\'enyi objective does not in general yield an unbiased objective estimator.

The vine structure and pair-copula families may be specified in advance, in which case only the tree-wise estimation procedure is required. When they are unknown, we adapt the sequential selection algorithm of \citet{dissmann2013selecting}. The original procedure constructs each tree through a maximum spanning-tree problem using edge weights based on absolute empirical Kendall's $\tau$ and subsequently selects a pair-copula family for every selected edge. In our approach, every admissible edge is fitted under each candidate pair-copula family and the resulting covariate-dependent pair-copula regressions are evaluated using an information criterion based on effective degrees of freedom. For each candidate edge, the family minimizing the chosen criterion is retained, and the negative information-criterion value is used as its edge weight in the maximum spanning-tree problem. The corresponding already fitted pair-copula regressions are retained after tree selection and provide the conditional pseudo-observations required at the next level.

The main contributions of this paper are threefold. First, we propose a truly multivariate structured additive distributional regression model based on regular vines, combining potentially heterogeneous marginal distributions with pair-specific copula families whose parameters may vary with covariates. Second, we develop a scalable tree-wise stochastic variational inference procedure based on component-specific Gaussian variational approximations. Third, we adapt sequential vine selection to the distributional-regression setting by selecting pair-copula families and weighting admissible edges through information criteria applied to fitted covariate-dependent pair-copula regressions.

The remainder of the paper is organized as follows. Section~\ref{sec:sadr} introduces structured additive distributional regression and its prior specification. Section~\ref{sec:vine_model} introduces regular vines and combines them with structured additive marginal and pair-copula regressions. Section~\ref{sec:svi} presents the variational approximation, tree-wise estimation, and the selection of the vine structure and pair-copula families.

\section{Marginal structured additive distributional regression}\label{sec:sadr}

\subsection{Model specification}\label{sec:sadr_model}

Let $\{(\Y_i,\x_i):i=1,\dots,n\}$ denote independent observations, where $\Y_i=(Y_{i,1},\dots,Y_{i,D})^\T$ is a $D$-dimensional response vector and $\x_i$ denotes the observed covariate vector. We first consider the conditional marginal distribution of a single response component $Y_{i,d}$, $d=1,\dots,D$. Conditionally on $\X_i=\x_i$, the marginal distribution of $Y_{i,d}$ is assumed to have density
\begin{align*}
p_d(y_{i,d}\mid\bs\nu_{i,d}),
\end{align*}
where $\bs\nu_{i,d}=(\nu_{i,d,1},\dots,\nu_{i,d,P_d})^\T$ contains the $P_d$ distributional parameters of margin $d$.
For every $p=1,\dots,P_d$, the parameter $\nu_{i,d,p}$ is related to a real-valued predictor through a suitable link function $g_{d,p}$,
\begin{align*}
g_{d,p}(\nu_{i,d,p})=\eta_{i,d,p},\qquad \nu_{i,d,p}=g_{d,p}^{-1}(\eta_{i,d,p}).
\end{align*}
The predictor is assumed to be of structured additive form,
\begin{align*}
\eta_{i,d,p}=\sum_{j=1}^{J_{d,p}}f_{d,p,j}(\x_{i,d,p,j},\bs\beta_{d,p,j}),
\end{align*}
where $f_{d,p,j}$ denotes the $j$th covariate effect entering parameter $p$ of margin $d$. The effects may be linear or smooth and are represented through basis expansions,
\begin{align*}
f_{d,p,j}(\x_{i,d,p,j},\bs\beta_{d,p,j})=\X_{i,d,p,j}\bs\beta_{d,p,j},
\end{align*}
where $\X_{i,d,p,j}$ is a row vector of basis evaluations and $\bs\beta_{d,p,j}\in\mathbb R^{Q_{d,p,j}}$ is the corresponding coefficient vector. Collecting the evaluations over all observations gives
\begin{align*}
\bs f_{d,p,j}=\X_{d,p,j}\bs\beta_{d,p,j}.
\end{align*}
The identifiability constraints imposed on the additive effects and the corresponding penalty reparameterization are described in \autoref{sec:spline_constraints}.

\subsection{Prior specification}\label{sec:sadr_prior}

Smooth effects are regularized through quadratic penalties. Let $\K_{d,p,j}$ denote the penalty matrix associated with effect $j$ of parameter $p$ in margin $d$, and let $\tau_{d,p,j}^2>0$ denote the corresponding smoothing variance. Conditional on $\tau_{d,p,j}^2$, the smoothing prior contribution is
\begin{align}
p(\bs\beta_{d,p,j}\mid\tau_{d,p,j}^2)
\propto
\exp\left\{
-\frac{1}{2\tau_{d,p,j}^2}
\bs\beta_{d,p,j}^\T
\K_{d,p,j}
\bs\beta_{d,p,j}
\right\}.
\label{eq:smoothing_prior}
\end{align}
Smaller values of $\tau_{d,p,j}^2$ induce stronger regularization, whereas larger values allow more variable fitted effects. Each smoothing variance is assigned a proper hyperprior,
\begin{align*}
\tau_{d,p,j}^2
\sim
p_{\tau,d,p,j}.
\end{align*}

The penalty matrix $\K_{d,p,j}$ may be rank deficient. As described in Appendix~\ref{sec:spline_constraints}, each smooth effect is therefore reparameterized into penalized and unpenalized coefficient blocks. Under this reparameterization, \eqref{eq:smoothing_prior} acts only on the penalized block and takes the form
\begin{align*}
p\left(
\bs\beta^{\mathrm{penalized}}_{d,p,j}
\mid
\tau_{d,p,j}^2
\right)
\propto
\exp\left\{
-\frac{1}{2\tau_{d,p,j}^2}
\left(\bs\beta^{\mathrm{penalized}}_{d,p,j}\right)^\T
\bs\beta^{\mathrm{penalized}}_{d,p,j}
\right\}.
\end{align*}
Intercepts and unpenalized coefficient blocks are left unpenalized and do not contribute a prior term. The same construction is used for the smooth effects entering the pair-copula regressions introduced in Section~\ref{sec:vine_model}.

\section{Truly multivariate vine copula distributional regression}\label{sec:vine_model}

We now extend the marginal distributional regression framework to the joint distribution of the $D$ response components. Their conditional marginal distributions are combined through a regular vine copula, which provides a flexible representation of the dependence structure.

\subsection{Copulas}\label{sec:copulas}

A $D$-dimensional copula is a cumulative distribution function on $[0,1]^D$ with standard uniform marginal distributions. Let $\Y=(Y_1,\dots,Y_D)^\T$ have continuous marginal distribution functions $F_1,\dots,F_D$ and joint distribution function $F$. By Sklar's theorem \citep{sklar1959fonctions}, there exists a unique copula $C$ such that
\begin{align*}
F(y_1,\dots,y_D)=C\left(F_1(y_1),\dots,F_D(y_D)\right).
\end{align*}
When all densities exist,
\begin{align*}
p(y_1,\dots,y_D)=c(u_1,\dots,u_D)\prod_{d=1}^D p_d(y_d),\qquad u_d=F_d(y_d),
\end{align*}
where $c$ denotes the density of $C$.

In regression settings, all distributions are conditional on the observed covariates. Provided that the same conditioning information is used for the margins and the copula, the conditional extension of Sklar's theorem gives \citep{patton2006modelling}
\begin{align*}
F(y_1,\dots,y_D\mid\x_i)=C\left(F_1(y_1\mid\x_i),\dots,F_D(y_D\mid\x_i);\bs\theta_i\right),\qquad \bs\theta_i\equiv\bs\theta(\x_i).
\end{align*}
Here, $\bs\theta_i$ contains the observation-specific dependence parameters. For continuous conditional marginals, $U_{i,d}=F_d(Y_{i,d}\mid\x_i)$ is uniformly distributed on $[0,1]$ conditionally on $\X_i=\x_i$.

Directly specifying a flexible $D$-dimensional copula becomes increasingly difficult as $D$ grows. Pair-copula constructions address this problem by decomposing a multivariate copula into a sequence of bivariate unconditional and conditional copulas \citep{bedford2001probability,bedford2002vines,aas2009,dissmann2013selecting,czado2022vine}.

\subsection{Regular vines}\label{sec:regular_vines}

A regular vine determines which unconditional and conditional pairs enter a pair-copula construction through a sequence of linked trees.

\begin{definition}[Regular vine]\label{def:regular_vine}
A regular vine on $D$ elements is a sequence $\mathcal V=(T_1,\dots,T_{D-1})$ of trees $T_m=(V_m,E_m)$ such that $V_1=\{1,\dots,D\}$, $V_m=E_{m-1}$ for $m=2,\dots,D-1$, and two nodes in $T_m$ may be joined only if the corresponding edges in $T_{m-1}$ share exactly one node. The latter requirement is referred to as the \emph{proximity condition}.
\end{definition}

Thus, the nodes of the first tree correspond to the original variables, while the nodes of every subsequent tree are the edges of the preceding tree. Following \citet{dissmann2013selecting}, an edge $e_m\in E_m$ is represented by the conditional pair $(j_{e_m},k_{e_m};D_{e_m})$, where $j_{e_m}$ and $k_{e_m}$ are the conditioned variables and $D_{e_m}$ is the conditioning set. The proximity condition implies
\begin{align*}
|D_{e_m}|=m-1.
\end{align*}
For $m=1$, $D_{e_1}=\emptyset$ and the corresponding pairs are unconditional. Since tree $T_m$ contains $D-m$ edges, a complete regular vine contains
\begin{align*}
\sum_{m=1}^{D-1}(D-m)=\frac{D(D-1)}{2}
\end{align*}
pair-copula components.

Associating one bivariate copula $C_{j_{e_m},k_{e_m};D_{e_m}}$ with every edge yields an R-vine copula. In general, a conditional pair copula may depend on the realized values of its conditioning variables. We write $C_{j_{e_m},k_{e_m};D_{e_m}}(u,v\mid\bs u_{D_{e_m}})$ for the conditional copula of $U_{j_{e_m}}$ and $U_{k_{e_m}}$ given $\bs U_{D_{e_m}}=\bs u_{D_{e_m}}$, with corresponding density $c_{j_{e_m},k_{e_m};D_{e_m}}$. For a conditioning set $A\subseteq\{1,\dots,D\}$ and $j\notin A$, define
\begin{align*}
u_{j\mid A}
=
F_{U_j\mid\bs U_A}\left(u_j\mid\bs u_A\right),
\end{align*}
with $u_{j\mid\emptyset}=u_j$. The resulting R-vine copula density is
\begin{align*}
c_{\mathcal V}(\bs u)
=
\prod_{m=1}^{D-1}\prod_{e_m\in E_m}
c_{j_{e_m},k_{e_m};D_{e_m}}
\left(
u_{j_{e_m}\mid D_{e_m}},
u_{k_{e_m}\mid D_{e_m}}
\mid
\bs u_{D_{e_m}}
\right).
\end{align*}

The conditional arguments entering higher vine trees are obtained recursively from the pair copulas in preceding trees. For a continuously differentiable conditional pair copula, define the $h$-functions
\begin{align*}
h_{j_{e_m}\mid k_{e_m};D_{e_m}}\left(u,v\mid\bs u_{D_{e_m}}\right)
&=
\frac{\partial}{\partial v}
C_{j_{e_m},k_{e_m};D_{e_m}}
\left(u,v\mid\bs u_{D_{e_m}}\right),\\
h_{k_{e_m}\mid j_{e_m};D_{e_m}}\left(v,u\mid\bs u_{D_{e_m}}\right)
&=
\frac{\partial}{\partial u}
C_{j_{e_m},k_{e_m};D_{e_m}}
\left(u,v\mid\bs u_{D_{e_m}}\right).
\end{align*}
The corresponding conditional probability integral transforms satisfy
\begin{align}
u_{j_{e_m}\mid D_{e_m}\cup\{k_{e_m}\}}
&=
h_{j_{e_m}\mid k_{e_m};D_{e_m}}
\left(
u_{j_{e_m}\mid D_{e_m}},
u_{k_{e_m}\mid D_{e_m}}
\mid
\bs u_{D_{e_m}}
\right),
\label{eq:h_recursion_first}\\
u_{k_{e_m}\mid D_{e_m}\cup\{j_{e_m}\}}
&=
h_{k_{e_m}\mid j_{e_m};D_{e_m}}
\left(
u_{k_{e_m}\mid D_{e_m}},
u_{j_{e_m}\mid D_{e_m}}
\mid
\bs u_{D_{e_m}}
\right).
\label{eq:h_recursion_second}
\end{align}
These recursions provide the conditional arguments required in subsequent vine trees.

\begin{figure}[t]
    \centering
    \includegraphics[width=0.75\textwidth]{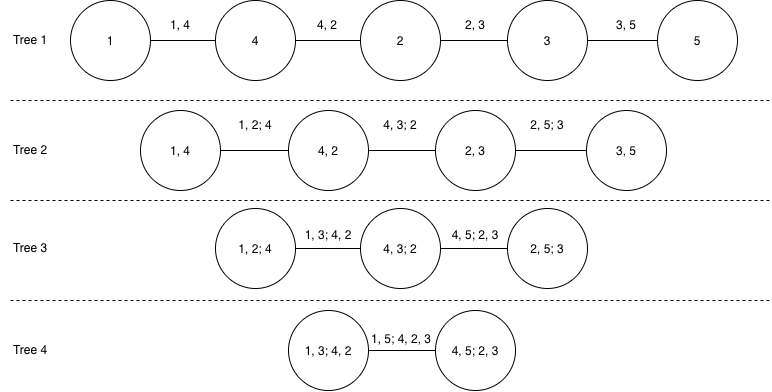}
    \caption{Graphical representation of the five-dimensional R-vine used in the simulation study. Nodes in $T_1$ correspond to response components, while nodes in subsequent trees correspond to edges of the preceding tree and are labelled by the associated conditioned variables and conditioning set.}
    \label{fig:rvine_graphical_representation}
\end{figure}

For storage and computation, a regular vine can equivalently be encoded by a triangular \emph{R-vine array} \citep{dissmann2013selecting,czado2022vine}. We use the upper-left triangular convention throughout this paper. For example, consider the five-dimensional R-vine array
\begin{align}
\M_{\mathrm{sim}}
=
\begin{pmatrix}
4&2&3&5&5\\
2&3&5&3&0\\
3&5&2&0&0\\
5&4&0&0&0\\
1&0&0&0&0
\end{pmatrix}.
\label{eq:rvine_array_example}
\end{align}
The corresponding edge sets are
\begin{align*}
T_1 &: \quad (1,4),\ (4,2),\ (2,3),\ (3,5), \\
T_2 &: \quad (1,2;4),\ (4,3;2),\ (2,5;3), \\
T_3 &: \quad (1,3;4,2),\ (4,5;2,3), \\
T_4 &: \quad (1,5;4,2,3).
\end{align*}
The same vine is displayed graphically in Figure~\ref{fig:rvine_graphical_representation}. In $T_1$, the nodes correspond to the response components and the edges to unconditional pair copulas. From $T_2$ onward, the nodes correspond to edges of the preceding tree, and the proximity condition determines which nodes may be connected.

Throughout the paper, we adopt the standard simplifying assumption, which removes the direct dependence of a conditional pair copula on the realized values of its conditioning variables and is widely used in vine-copula modelling \citep{stoeber2013,czado2022vine,nagler2025simplified}. Specifically, for every $e_m\in E_m$,
\begin{align*}
C_{j_{e_m},k_{e_m};D_{e_m}}
\left(
u,v\mid\bs U_{D_{e_m}}=\bs u_{D_{e_m}}
\right)
=
C_{j_{e_m},k_{e_m};D_{e_m}}(u,v).
\end{align*}
Under this assumption, each conditional pair copula is represented by an ordinary bivariate copula whose parameters do not depend on the realized conditioning variables.

\subsection{Vine copula distributional regression}\label{sec:vine_distributional_regression}

We combine the simplified R-vine construction with structured additive distributional regression by allowing the parameters of both the marginal distributions and the pair copulas to depend on observed covariates. In particular, although the simplifying assumption removes direct dependence of a pair copula on the realized conditioning variables, its parameters may still vary with external covariates, as in covariate-dependent pair-copula constructions \citep{vatter2018generalized}. Thus,
\begin{align*}
C_{j_{e_m},k_{e_m};D_{e_m}}
\left(
u,v
\mid
\bs U_{D_{e_m}}=\bs u_{D_{e_m}},
\X_i=\x_i
\right)
=
C_{j_{e_m},k_{e_m};D_{e_m}}
\left(
u,v;
\bs\theta_{i,e_m}
\right),
\end{align*}
where $\bs\theta_{i,e_m}\equiv\bs\theta_{e_m}(\x_i)$. Accordingly, under the simplifying assumption the associated $h$-functions depend on the covariates through $\bs\theta_{i,e_m}$, and we write them as $h_{j_{e_m}\mid k_{e_m};D_{e_m}}(u,v;\bs\theta_{i,e_m})$ and $h_{k_{e_m}\mid j_{e_m};D_{e_m}}(v,u;\bs\theta_{i,e_m})$. For observation $i$, let
\begin{align*}
u_{i,d}=F_d(y_{i,d}\mid\bs\nu_{i,d}),\qquad d=1,\dots,D,
\end{align*}
where the components of $\bs\nu_{i,d}$ are related to $\x_i$ through the structured additive predictors introduced in Section~\ref{sec:sadr_model}. For an edge $e_m$, write $u_{i,j_{e_m}\mid D_{e_m}}$ and $u_{i,k_{e_m}\mid D_{e_m}}$ for the corresponding conditional probability integral transforms obtained recursively through \eqref{eq:h_recursion_first}--\eqref{eq:h_recursion_second}. All conditional distributions involved in these transformations are additionally conditional on $\X_i=\x_i$, which is suppressed from the notation.

Combining the marginal distributional regressions with the R-vine copula gives
\begin{align}
p(\Y_i=\y_i\mid\X_i=\x_i)
&=
\prod_{d=1}^D p_d(y_{i,d}\mid\bs\nu_{i,d})
\nonumber\\
&\quad\times
\prod_{m=1}^{D-1}\prod_{e_m\in E_m}
c_{j_{e_m},k_{e_m};D_{e_m}}
\left(
u_{i,j_{e_m}\mid D_{e_m}},
u_{i,k_{e_m}\mid D_{e_m}};
\bs\theta_{i,e_m}
\right).
\label{eq:full_vine_regression_density}
\end{align}
The first product specifies the conditional marginal distributions of the individual response components, while the second specifies their conditional dependence.

The marginal parameters $\bs\nu_{i,d}$ follow the structured additive specification in Section~\ref{sec:sadr_model}. Analogously, let pair copula $e_m$ have $P_{e_m}$ parameters $\bs\theta_{i,e_m}=(\theta_{i,e_m,1},\dots,\theta_{i,e_m,P_{e_m}})^\T$. For $p=1,\dots,P_{e_m}$,
\begin{align}
g_{e_m,p}(\theta_{i,e_m,p})&=\eta_{i,e_m,p},\qquad \theta_{i,e_m,p}=g_{e_m,p}^{-1}(\eta_{i,e_m,p}), \nonumber\\
\eta_{i,e_m,p}&=\sum_{j=1}^{J_{e_m,p}}f_{e_m,p,j}(\x_{i,e_m,p,j},\bs\beta_{e_m,p,j}).
\label{eq:pair_copula_predictor}
\end{align}
The basis representations, smoothing priors, and identifiability constraints introduced for the marginal distributional regressions carry over directly, with the edge index $e_m$ replacing the margin index $d$.

The model in \eqref{eq:full_vine_regression_density} is truly multivariate because it is defined for arbitrary $D\geq2$, permits different marginal distributions across response components, and provides dependence flexibility at two distinct levels. Across vine edges, different pair-copula families may be used, allowing asymmetric, tail-dependent, and tail-independent forms of association to coexist within the same multivariate model. Within each edge, every parameter of the selected pair copula may itself vary with covariates through a structured additive predictor. Consequently, both the qualitative form of dependence across response pairs and the strength and shape of dependence within a given pair may vary flexibly across the covariate space. This provides substantially greater dependence flexibility than Gaussian-copula structured additive distributional regression \citep{kock2023truly}.

\section{Stochastic variational inference and tree-wise estimation}\label{sec:svi}

\subsection{Componentwise variational approximation}\label{sec:variational_approximation}

Let $\mathcal M=\{1,\dots,D\}$ denote the set of marginal regression components and let
\begin{align*}
\mathcal E=\bigcup_{m=1}^{D-1}E_m
\end{align*}
denote the set of pair-copula regression components associated with the regular vine $\mathcal V$. We write
\begin{align*}
\mathcal R=\mathcal M\cup\mathcal E
\end{align*}
for the set of all regression components of the model, such that $\mathcal M\subset\mathcal R$ and $\mathcal E\subset\mathcal R$. An element $r=d\in\mathcal M$ corresponds to one marginal distributional regression, whereas $r=e_m\in\mathcal E$ corresponds to one pair-copula distributional regression.

For margin $d$, let $\bs\beta_d$ collect the coefficient vectors $\bs\beta_{d,p,j}$ introduced in Section~\ref{sec:sadr}, for $p=1,\dots,P_d$ and $j=1,\dots,J_{d,p}$. For every penalized effect, let $\tau_{d,p,j}^2>0$ denote the smoothing variance introduced in Section~\ref{sec:sadr} and define
\begin{align*}
\widetilde{\tau}_{d,p,j}
=
\log \tau_{d,p,j}^2.
\end{align*}
Analogously, for a pair-copula component $e_m$, let $\bs\beta_{e_m}$ collect all regression coefficients and define
\begin{align*}
\widetilde{\tau}_{e_m,p,j}
=
\log \tau_{e_m,p,j}^2
\end{align*}
for every penalized pair-copula effect.

For every component $r\in\mathcal R$, the regression coefficients are represented jointly by a multivariate Gaussian variational distribution,
\begin{align*}
q_{\bs\phi_{\beta,r}}(\bs\beta_r)
=
\mathcal N
\left(
\bs\beta_r
\mid
\bs\mu_{\beta,r},
\bs L_{\beta,r}\bs L_{\beta,r}^{\T}
\right),
\end{align*}
where $\bs L_{\beta,r}$ is lower triangular with positive diagonal entries. Thus, dependence between regression coefficients belonging to different distributional parameters and effects of the same marginal or pair-copula regression is retained through a full covariance matrix.

Each log-smoothing variance is assigned a separate univariate Gaussian variational factor,
\begin{align*}
q_{\bs\phi_{\tau,r}}
\left(
\{\widetilde{\tau}_{r,k}\}_{k=1}^{S_r}
\right)
=
\prod_{k=1}^{S_r}
\mathcal N
\left(
\widetilde{\tau}_{r,k}
\mid
\mu_{\tau,r,k},
\sigma_{\tau,r,k}^2
\right),
\end{align*}
where $S_r$ denotes the number of smoothing variances in component $r$. The corresponding smoothing variances are obtained through
\begin{align*}
\tau_{r,k}^2
=
\exp\left(\widetilde{\tau}_{r,k}\right).
\end{align*}

The component-specific variational approximation is therefore
\begin{align*}
q_{\bs\phi_r}
\left(
\bs\beta_r,
\{\widetilde{\tau}_{r,k}\}_{k=1}^{S_r}
\right)
=
q_{\bs\phi_{\beta,r}}(\bs\beta_r)
\prod_{k=1}^{S_r}
q_{\phi_{\tau,r,k}}
\left(
\widetilde{\tau}_{r,k}
\right).
\end{align*}

Let $\bs\phi=\{\bs\phi_r:r\in\mathcal R\}$ collect the variational parameters of all model components. The complete variational approximation is
\begin{align}
q_{\bs\phi}
\left(
\{\bs\beta_r,\widetilde{\bs\tau}_r\}_{r\in\mathcal R}
\right)
=
\prod_{r\in\mathcal R}
q_{\bs\phi_r}
\left(
\bs\beta_r,\widetilde{\bs\tau}_r
\right).
\label{eq:joint_variational_factorization}
\end{align}
Thus, the variational family factorizes across marginal and pair-copula components, between regression coefficients and smoothing variances within each component, and across the individual log-smoothing variances. Full covariance is retained within the regression-coefficient block of each component. The tree-wise and joint estimation strategies considered below use the same variational family and differ only in how its component-specific parameters are optimized.

\subsection{Evidence lower bound and mini-batching}\label{sec:elbo}

For component $r\in\mathcal R$, let $\ell_{i,r}(\bs\beta_r)$ denote the contribution of observation $i$ to its log-likelihood. For a marginal component $r=d$,
\begin{align*}
\ell_{i,d}(\bs\beta_d)
=
\log p_d(y_{i,d}\mid\bs\nu_{i,d}),
\end{align*}
whereas for a pair-copula component $r=e_m$ it is the log pair-copula density evaluated at the conditional probability integral transforms associated with that edge.

We denote by $\log p_r(\bs\beta_r,\widetilde{\bs\tau}_r)$ the contribution of the smoothing priors and hyperpriors expressed in terms of the log-smoothing variances.

The variational parameters $\bs\phi_r=(\bs\phi_{\beta,r},\bs\phi_{\tau,r})$ are estimated by maximizing the componentwise evidence lower bound
\begin{align}
\mathcal L_r(\bs\phi_r)
=
\mathbb E_{q_{\bs\phi_r}}
\left[
\sum_{i=1}^n
\ell_{i,r}(\bs\beta_r)
+
\log p_r
\left(
\bs\beta_r,\widetilde{\bs\tau}_r
\right)
-
\log q_{\bs\phi_r}
\left(
\bs\beta_r,\widetilde{\bs\tau}_r
\right)
\right].
\label{eq:component_elbo}
\end{align}

For a uniformly sampled mini-batch $\mathcal B\subset\{1,\dots,n\}$ of size $B$, the full-data log-likelihood contribution is replaced by
\begin{align}
\widehat{\ell}_{\mathcal B,r}(\bs\beta_r)
=
\frac{n}{B}
\sum_{i\in\mathcal B}
\ell_{i,r}(\bs\beta_r).
\label{eq:minibatch_likelihood}
\end{align}
Since
\begin{align*}
\E_{\mathcal B}
\left[
\widehat{\ell}_{\mathcal B,r}(\bs\beta_r)
\right]
=
\sum_{i=1}^n
\ell_{i,r}(\bs\beta_r),
\end{align*}
this yields an unbiased stochastic estimator of the likelihood contribution to \eqref{eq:component_elbo}. The resulting ELBO is optimized using standard Monte Carlo reparameterization and stochastic gradient optimization.

Mini-batching is particularly useful in the present setting. For a fixed complete vine, the model contains $D$ marginal regressions and $D(D-1)/2$ pair-copula regressions. If the vine structure and pair-copula families are unknown, additional candidate regressions must be fitted when constructing each tree.

We use the standard ELBO rather than the R\'enyi objective considered by \citet{griesbauer2026stepwise}. Their setting concerns a vine copula used within the variational posterior, whereas the vine in the present model enters the response likelihood. Moreover, although the rescaled observation-level mini-batch log-likelihood in \eqref{eq:minibatch_likelihood} is unbiased, substituting it into the nonlinear R\'enyi objective does not in general yield an unbiased objective estimator.

\subsection{Tree-wise estimation for a fixed vine}\label{sec:treewise_estimation}

We first consider a known regular-vine structure $\mathcal V=(T_1,\dots,T_{D-1})$ with fixed pair-copula families. The marginal distributional regressions are estimated separately by maximizing \eqref{eq:component_elbo}. Throughout the sequential construction, fitted quantities are obtained by evaluating the corresponding structured additive predictors at the variational posterior mean of the regression coefficients. Specifically, for a fitted component $r\in\mathcal R$, we define
\begin{align*}
\widehat{\bs\beta}_r
=
\mathbb E_{q_{\bs\phi_{\beta,r}}}
\left[
\bs\beta_r
\right]
=
\bs\mu_{\beta,r}.
\end{align*}
Hatted distributional parameters denote the corresponding plug-in values obtained from $\widehat{\bs\beta}_r$ through the appropriate predictors and inverse link functions. Thus, $\widehat{\bs\nu}_{i,d}$ and $\widehat{\bs\theta}_{i,e_m}$ are fitted parameter vectors evaluated at $\widehat{\bs\beta}_d$ and $\widehat{\bs\beta}_{e_m}$, respectively; they do not in general coincide with posterior means on the distributional-parameter scale.

The fitted conditional distribution functions then provide the marginal probability integral transforms
\begin{align}
\widehat u_{i,d}
=
F_d\left(y_{i,d}\mid\widehat{\bs\nu}_{i,d}\right),
\qquad
i=1,\dots,n,\quad d=1,\dots,D.
\label{eq:fitted_marginal_pits}
\end{align}

The pair-copula regressions are subsequently estimated along the vine trees. Consider an edge $e_m\in E_m$ with conditioned variables $j_{e_m}$ and $k_{e_m}$ and conditioning set $D_{e_m}$, using the notation of Section~\ref{sec:regular_vines}. Its observation-specific log-likelihood contribution is
\begin{align*}
\ell_{i,e_m}(\bs\beta_{e_m})
=
\log
c_{j_{e_m},k_{e_m};D_{e_m}}
\left(
\widehat u_{i,j_{e_m}\mid D_{e_m}},
\widehat u_{i,k_{e_m}\mid D_{e_m}};
\bs\theta_{i,e_m}
\right),
\end{align*}
where $\bs\theta_{i,e_m}$ is obtained from the structured additive predictors in \eqref{eq:pair_copula_predictor}. The conditional probability integral transforms entering this likelihood contribution are treated as fixed while the component is estimated.

Once the pair-copula regressions in tree $T_m$ have been estimated, their fitted $h$-functions are used to construct the conditional pseudo-observations required by the subsequent tree. For $e_m\in E_m$,
\begin{align}
\widehat u_{i,j_{e_m}\mid D_{e_m}\cup\{k_{e_m}\}}
&=
h_{j_{e_m}\mid k_{e_m};D_{e_m}}
\left(
\widehat u_{i,j_{e_m}\mid D_{e_m}},
\widehat u_{i,k_{e_m}\mid D_{e_m}};
\widehat{\bs\theta}_{i,e_m}
\right),
\label{eq:fitted_h_first}\\
\widehat u_{i,k_{e_m}\mid D_{e_m}\cup\{j_{e_m}\}}
&=
h_{k_{e_m}\mid j_{e_m};D_{e_m}}
\left(
\widehat u_{i,k_{e_m}\mid D_{e_m}},
\widehat u_{i,j_{e_m}\mid D_{e_m}};
\widehat{\bs\theta}_{i,e_m}
\right).
\label{eq:fitted_h_second}
\end{align}
The fitted margins and pair copulas from preceding trees remain fixed when estimating later trees. Thus, estimation proceeds from the marginal regressions through $T_1,\dots,T_{D-1}$, with \eqref{eq:fitted_h_first}--\eqref{eq:fitted_h_second} recursively providing the inputs required at each subsequent level.

\subsection{Interpretation of the sequential approximation}\label{sec:sequential_interpretation}

The tree-wise procedure does not coincide with joint inference under the complete likelihood in \eqref{eq:full_vine_regression_density}. The marginal probability integral transforms entering $T_1$ are computed from fitted marginal regressions and subsequently treated as fixed. Likewise, the conditional pseudo-observations entering tree $T_m$ are computed from the fitted pair copulas in preceding trees and are treated as fixed when estimating the components of $T_m$.

The variational approximation therefore represents uncertainty within each marginal or pair-copula regression, but does not propagate uncertainty from previously estimated components through the recursive $h$-function construction. Posterior summaries for a pair-copula regression are conditional on the fitted margins, the preceding pair-copula regressions, and the resulting conditional pseudo-observations. This sequential approximation reduces the full multivariate estimation problem to a sequence of marginal and bivariate distributional regressions.
\subsection{Vine structure and pair-copula family selection}\label{sec:vine_selection}

When the vine structure or pair-copula families are unknown, we construct the vine sequentially by adapting the maximum spanning-tree strategy of \citet{dissmann2013selecting}. The original procedure uses absolute empirical Kendall's $\tau$ as the edge weight when constructing each tree and subsequently selects a pair-copula family for every selected edge. In the present setting, however, the pair-copula parameters vary with covariates, so a single unconditional dependence measure does not directly reflect the fitted conditional dependence structure. We therefore fit covariate-dependent pair-copula regressions for all admissible candidate edges and use an information criterion both to select the pair-copula family and to determine the edge weights entering the maximum spanning-tree problem.

Suppose trees $\widehat T_1,\dots,\widehat T_{m-1}$ have already been selected. Let $\mathcal C_m$ denote the set of admissible candidate edges for tree $T_m$. For $m=1$, $\mathcal C_1$ contains all pairs of marginal indices. For $m\geq2$, candidates are pairs of nodes in $\widehat T_{m-1}$ satisfying the proximity condition in Definition~\ref{def:regular_vine}. Their conditional pseudo-observations are those obtained from the pair-copula fits retained in the preceding trees.

Let $\mathfrak C$ denote the set of candidate pair-copula families. We regard the independence copula as a parameter-free pair-copula regression with an empty coefficient vector. For every candidate edge $e_m\in\mathcal C_m$ and every family $c\in\mathfrak C$, we fit the corresponding pair-copula distributional regression. Following the plug-in convention introduced above, fitted quantities are evaluated at the variational posterior mean of the regression coefficients. The fitted conditional log-likelihood is
\begin{align}
\widehat{\ell}_{e_m,c}
=
\sum_{i=1}^n
\log
c^{(c)}_{j_{e_m},k_{e_m};D_{e_m}}
\left(
\widehat u_{i,j_{e_m}\mid D_{e_m}},
\widehat u_{i,k_{e_m}\mid D_{e_m}};
\widehat{\bs\theta}_{i,e_m,c}
\right).
\label{eq:candidate_loglik}
\end{align}

To account for differences in model complexity across candidate pair-copula regressions, we use effective degrees of freedom. Let $\bs H_{e_m,c}$ denote the observed negative Hessian of the unpenalized conditional log-likelihood with respect to the regression coefficients, evaluated at the fitted coefficient vector, and let $\widehat{\bs K}_{e_m,c}$ denote the corresponding block-diagonal penalty matrix evaluated at the fitted smoothing variances. The latter contains zero blocks for intercepts and other unpenalized coefficients. For quadratically penalized regression models, the effective degrees of freedom can be expressed as \citep{wood2016smoothing}
\begin{align*}
\operatorname{df}_{e_m,c}
=
\operatorname{tr}
\left\{
\left(
\bs H_{e_m,c}
+
\widehat{\bs K}_{e_m,c}
\right)^{-1}
\bs H_{e_m,c}
\right\}.
\end{align*}

Let $\widehat{\bs\Sigma}_{e_m,c}$ denote the covariance matrix of the fitted variational distribution for the regression coefficients. We adapt the preceding expression to the variational approximation by using $\widehat{\bs\Sigma}_{e_m,c}$ as an approximation to the inverse penalized curvature,
\begin{align*}
\widehat{\bs\Sigma}_{e_m,c}
\approx
\left(
\bs H_{e_m,c}
+
\widehat{\bs K}_{e_m,c}
\right)^{-1}.
\end{align*}
Consequently,
\begin{align*}
\widehat{\operatorname{df}}_{e_m,c}
&=
Q_{e_m,c}
-
\operatorname{tr}
\left(
\widehat{\bs\Sigma}_{e_m,c}
\widehat{\bs K}_{e_m,c}
\right),
\end{align*}
where $Q_{e_m,c}$ is the dimension of the regression-coefficient vector. The smoothing variances enter through $\widehat{\bs K}_{e_m,c}$ and are treated as fixed at their fitted values when computing the effective degrees of freedom. For the parameter-free independence copula, $Q_{e_m,c}=0$, so the effective degrees of freedom are zero.

We consider the AIC
\begin{align*}
\operatorname{AIC}_{e_m,c}
=
-2\widehat{\ell}_{e_m,c}
+
2\widehat{\operatorname{df}}_{e_m,c}
\end{align*}
and the BIC-type criterion
\begin{align*}
\operatorname{BIC}^{\mathrm{edf}}_{e_m,c}
=
-2\widehat{\ell}_{e_m,c}
+
\log(n)\widehat{\operatorname{df}}_{e_m,c}.
\end{align*}
The latter is referred to as BIC-type because the usual parameter dimension is replaced by the effective degrees of freedom.

Let $\operatorname{IC}_{e_m,c}$ denote the information criterion chosen for a given vine fit, either $\operatorname{AIC}_{e_m,c}$ or $\operatorname{BIC}^{\mathrm{edf}}_{e_m,c}$. For each candidate edge, the selected pair-copula family is
\begin{align}
\widehat c_{e_m}
=
\operatorname*{arg\,min}_{c\in\mathfrak C}
\operatorname{IC}_{e_m,c},
\label{eq:candidate_family_selection}
\end{align}
and its edge weight is defined as
\begin{align}
w_{e_m}
=
-\operatorname{IC}_{e_m,\widehat c_{e_m}}
=
-\min_{c\in\mathfrak C}
\operatorname{IC}_{e_m,c}.
\label{eq:candidate_edge_weight}
\end{align}
Thus, the same criterion is used both to select the pair-copula family for each candidate edge and to score that edge for the subsequent tree selection.

Let $\mathfrak T_m$ denote the set of admissible spanning trees on the current node set. The selected tree is
\begin{align}
\widehat T_m
=
\operatorname*{arg\,max}_{T\in\mathfrak T_m}
\sum_{e_m\in E(T)}
w_{e_m}.
\label{eq:vine_mst}
\end{align}
Equivalently, because $w_{e_m}$ is the negative information criterion of the selected family, the maximum spanning tree minimizes the sum of the selected edgewise information criteria. For every edge $e_m\in E(\widehat T_m)$, we retain the already fitted regression corresponding to $\widehat c_{e_m}$. No additional pair-copula fit is performed after solving the maximum spanning-tree problem. The $h$-functions of these retained fits generate the conditional pseudo-observations used to form the admissible candidates for tree $T_{m+1}$. The procedure continues until $T_{D-1}$ has been selected.

\section{Simulation study}\label{sec:simulation}

We consider a five-dimensional response with sample size $n=2000$ and $R=100$ independent simulation replications. For replication $r=1,\dots,R$, the scalar covariates are independently sampled as $x_{r,i}\sim\mathcal U(-\pi,\pi)$, $i=1,\dots,n$. In the specification of the data-generating process below, we suppress the replication index and write $x_i$ and $Y_{i,d}$ for notational simplicity. The regular-vine structure, marginal distributions, and pair-copula families are treated as known throughout the experiment.

The simulation study has three main goals. First, we assess whether the proposed stepwise estimator can accurately recover nonlinear covariate effects in both the marginal and pair-copula regressions. Second, by comparing it with an oracle stepwise estimator based on the true recursive conditional probability integral transforms, we isolate the additional error introduced by using fitted pseudo-observations throughout the vine. Third, we compare the stepwise procedure with a joint global refinement in order to assess how propagating information through the complete vine likelihood affects point estimation and uncertainty quantification. By treating the vine structure and pair-copula families as known, the experiment focuses specifically on the inferential properties of the estimation procedures rather than on structure or family selection.

The scalar covariate $x_i$ enters every distributional parameter through one nonlinear effect. Normal and Gumbel distributions are parameterized in terms of location and scale, the exponential distribution in terms of its rate, and the Gamma distribution in terms of its mean and variance. The true marginal models are
\begin{align*}
Y_{i,1}\mid x_i &\sim \operatorname{Normal}\left(0.67 - 1.1\sin(x_i), \operatorname{softplus}\left(0.04 - 1.4\cos(x_i)\right)\right),\\
Y_{i,2}\mid x_i &\sim \operatorname{Gumbel}\left(-0.29 - 0.5\sin(x_i), \operatorname{softplus}\left(-0.25 + 0.8\sin(x_i)\right)\right),\\
Y_{i,3}\mid x_i &\sim \operatorname{Exponential}\left(\operatorname{softplus}\left(0.40 + 1.8\sin(x_i)\right)\right),\\
Y_{i,4}\mid x_i &\sim \operatorname{Gamma}\left(\operatorname{softplus}\left(0.52 + 0.9\sin(x_i)\right), \operatorname{softplus}\left(0.76 + 0.9\sin(x_i)\right)\right),\\
Y_{i,5}\mid x_i &\sim \operatorname{Normal}\left(0.75 - 0.7\cos(x_i), \operatorname{softplus}\left(0.53 - 1.3\sin(x_i)\right)\right).
\end{align*}

Dependence is described by the five-dimensional regular vine introduced in Section~\ref{sec:vine_model} and displayed graphically in Figure~\ref{fig:rvine_graphical_representation}. Its edge sets are
\begin{align*}
T_1 &: \quad (1,4),\ (4,2),\ (2,3),\ (3,5),\\
T_2 &: \quad (1,2;4),\ (4,3;2),\ (2,5;3),\\
T_3 &: \quad (1,3;4,2),\ (4,5;2,3),\\
T_4 &: \quad (1,5;4,2,3).
\end{align*}
Within each tree, the pair copulas are listed in the same order as the corresponding edges above. Their families and covariate-dependent parameter functions are
\begin{align*}
T_1:\quad&
\operatorname{Gumbel}
\left(
\cdot,\cdot;
1+\operatorname{softplus}(0.01+0.8\sin x_i)
\right),
\\
&
\operatorname{Gaussian}
\left(
\cdot,\cdot;
\tanh(0.63+0.8\sin x_i)
\right),
\\
&
\operatorname{Clayton}
\left(
\cdot,\cdot;
\operatorname{softplus}(0.32+1.3\cos x_i)
\right),
\\
&
\operatorname{Frank}
\left(
\cdot,\cdot;
0.38-0.9\cos x_i
\right),
\\
T_2:\quad&
\operatorname{Joe}
\left(
\cdot,\cdot;
1+\operatorname{softplus}(0.83-1.6\cos x_i)
\right),
\\
&
\operatorname{Gumbel}
\left(
\cdot,\cdot;
1+\operatorname{softplus}(-0.15+0.7\cos x_i)
\right),
\\
&
\operatorname{Gaussian}
\left(
\cdot,\cdot;
\tanh(0.79-0.8\sin x_i)
\right),
\\
T_3:\quad&
\operatorname{Clayton}
\left(
\cdot,\cdot;
\operatorname{softplus}(-0.28+1.6\sin x_i)
\right),
\\
&
\operatorname{Frank}
\left(
\cdot,\cdot;
-0.22+1.3\cos x_i
\right),
\\
T_4:\quad&
\operatorname{Gumbel}
\left(
\cdot,\cdot;
1+\operatorname{softplus}(0.26-1.7\cos x_i)
\right).
\end{align*}

The data-generating process therefore combines heterogeneous marginal distributions with Gaussian, Frank, Clayton, Gumbel, and Joe pair copulas, including symmetric, asymmetric, tail-dependent, and tail-independent dependence structures. In estimation, each nonlinear effect is represented by a spline with $10$ coefficients and regularized using the smoothing prior specified in Section~\ref{sec:sadr_prior}. For all smooth effects, the smoothing variances are assigned
\begin{align*}
\tau_{r,j}^2 \sim \operatorname{Inverse\text{-}Gamma}(1,1),
\end{align*}
with concentration and scale parameters both equal to $1$. The centering constraint and penalty reparameterization are implemented as described in Appendix~\ref{sec:spline_constraints}.

All models in the simulation study are implemented and estimated using the probabilistic programming framework \texttt{liesel} \citep{riebl2026lieselpythonframeworkgraphbased}, using the development branch \texttt{optima-ma-loss}.

\subsection{Estimation strategies}\label{sec:simulation_estimators}

We compare three estimation strategies.

\paragraph*{Stepwise estimation}

The stepwise estimator (SW) is the proposed procedure described in Section~\ref{sec:treewise_estimation}. The marginal distributional regressions are estimated first and their fitted conditional distribution functions are used to construct the marginal pseudo-observations. The pair-copula regressions are then estimated tree by tree. After estimating tree $T_m$, the fitted $h$-functions are used to construct the conditional pseudo-observations required by tree $T_{m+1}$.

\paragraph*{Oracle stepwise estimation}

The oracle estimator retains the same tree-wise estimation strategy but removes the error introduced by recursively estimated inputs. When a pair-copula regression is estimated, its conditional pseudo-observations are computed from the true marginal and pair-copula parameter functions of the data-generating process. Only the inputs to the current regression are therefore replaced by their oracle counterparts; the parameters of the current pair-copula regression are still estimated from the simulated data using the same variational approximation as under SW.

The comparison between SW and Oracle isolates the effect of using estimated pseudo-observations in the recursive vine construction. Since the current pair-copula regression is estimated in the same way under both procedures, differences between them arise from replacing the true marginal and conditional probability integral transforms by those obtained from the fitted margins and preceding pair copulas.

\paragraph*{Global refinement}

The third strategy first computes the complete SW solution and subsequently uses its variational parameters to initialize a joint optimization of the ELBO under the complete likelihood in \eqref{eq:full_vine_regression_density}. We refer to this estimator as Global. The joint variational approximation is the factorized distribution in \eqref{eq:joint_variational_factorization}; during global optimization, all component-specific variational parameters $\{\bs\phi_r:r\in\mathcal R\}$ are updated simultaneously. In particular, the conditional pseudo-observations are recomputed from the current upstream parameter values whenever the complete likelihood is evaluated. The Global estimator therefore does not treat the pseudo-observations produced by the initial SW fit as fixed.

The variational family itself is unchanged. SW and Global use the same component-specific Gaussian factors and the same factorization across model components. They differ in the objective used to optimize these factors: SW maximizes the componentwise ELBOs sequentially, whereas Global optimizes all factors jointly under the complete likelihood.

\subsection{Evaluation criteria}\label{sec:simulation_metrics}

We assess the recovery of the nonlinear covariate effects in terms of pointwise coverage, credible-interval width, and root mean squared error. All quantities are evaluated at the covariate values observed in the corresponding simulation replication and on the centered additive-effect scale.

Let $\mathcal P_\ell$ denote the collection of nonlinear effects belonging to level $\ell$, where $\ell=\mathrm{marg}$ denotes the marginal regressions and $\ell=T_m$ the pair-copula regressions in tree $T_m$. For $p\in\mathcal P_\ell$, let $f_p$ denote the corresponding uncentered true nonlinear effect. Since the additive effects are centered over the observed covariate values, the true centered effect in replication $r$ is
\begin{align*}
f_{r,p}(x_{r,i})
=
f_p(x_{r,i})
-
\frac{1}{n}
\sum_{s=1}^n
f_p(x_{r,s}),
\qquad i=1,\dots,n.
\end{align*}

For replication $r$, the mean pointwise coverage is
\begin{align*}
\operatorname{MPC}_{r,\ell}
=
\frac{1}{|\mathcal P_\ell|n}
\sum_{p\in\mathcal P_\ell}
\sum_{i=1}^n
\mathds 1
\left\{
f_{r,p}(x_{r,i})
\in
I_{r,p}^{0.95}(x_{r,i})
\right\},
\end{align*}
where $I_{r,p}^{0.95}(x_{r,i})$ denotes the pointwise $95\%$ credible interval for effect $p$ at the $i$th observed covariate value of replication $r$.

The mean interval width is
\begin{align*}
\operatorname{MW}_{r,\ell}
=
\frac{1}{|\mathcal P_\ell|n}
\sum_{p\in\mathcal P_\ell}
\sum_{i=1}^n
\left[
U_{r,p}(x_{r,i})
-
L_{r,p}(x_{r,i})
\right],
\end{align*}
where $L_{r,p}$ and $U_{r,p}$ denote the lower and upper credible-interval bounds.

Finally, point-estimation accuracy is measured by
\begin{align*}
\operatorname{RMSE}_{r,\ell}
=
\frac{1}{|\mathcal P_\ell|}
\sum_{p\in\mathcal P_\ell}
\left[
\frac{1}{n}
\sum_{i=1}^n
\left\{
\widehat f_{r,p}(x_{r,i})
-
f_{r,p}(x_{r,i})
\right\}^2
\right]^{1/2},
\end{align*}
where $\widehat f_{r,p}$ denotes the posterior mean estimate of the centered effect $f_{r,p}$. We compare the distributions of the three criteria across the $R=100$ simulation replications.

\subsection{Results}\label{sec:simulation_results}

\begin{figure}[t]
    \centering
    \includegraphics[width=\textwidth]{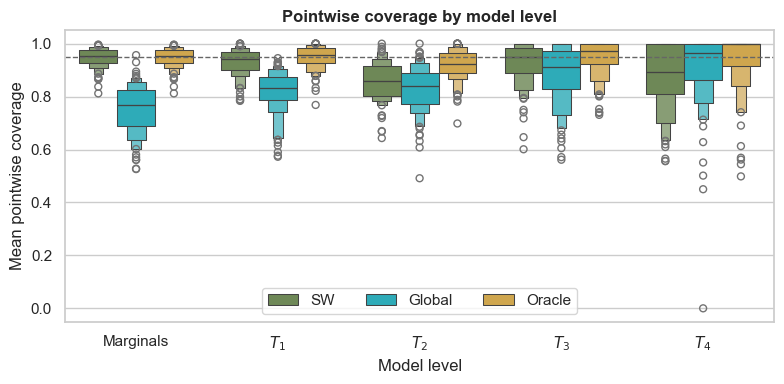}
    \caption{Mean pointwise coverage across the $100$ simulation replications for the stepwise (SW), globally refined (Global), and oracle stepwise (Oracle) estimators.}
    \label{fig:simulation_coverage}
\end{figure}

Figure~\ref{fig:simulation_coverage} reports the mean pointwise coverage for the three estimation strategies. SW and Oracle yield very similar coverage distributions over most model levels. In particular, the differences are small in the early vine trees, while some deterioration of SW relative to Oracle becomes visible at greater tree depths. This behaviour is consistent with the recursive construction of the conditional pseudo-observations: errors introduced at one level can enter the arguments of the pair copulas in all subsequent levels. Nevertheless, the overall proximity of SW and Oracle indicates that this propagation introduces only a moderate loss in calibration.

Global refinement produces a markedly different pattern. Coverage is lower than under SW and Oracle for the marginal regressions and the early vine trees, while the discrepancy decreases with tree depth. By the final tree, the three procedures give much more similar results.

\begin{figure}[t]
    \centering
    \includegraphics[width=\textwidth]{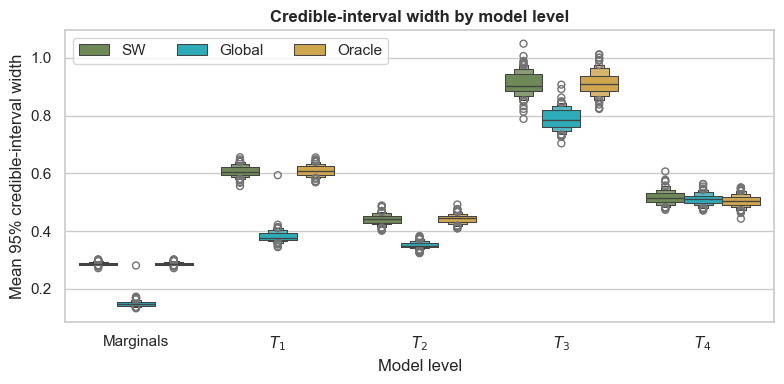}
    \caption{Mean pointwise width of the $95\%$ credible intervals across the $100$ simulation replications for the three estimation strategies.}
    \label{fig:simulation_width}
\end{figure}

The interval widths in Figure~\ref{fig:simulation_width} explain this difference in coverage. SW and Oracle produce almost indistinguishable credible-interval widths throughout the vine. Global refinement, by contrast, substantially reduces the interval widths for the marginal regressions and the first vine trees. The difference progressively decreases at higher tree levels and is small in the final tree.

This pattern reflects the recursive structure of the complete likelihood. Under SW, the parameters of a component are estimated from its local likelihood contribution conditional on the fitted quantities obtained in preceding stages. Under Global, an upstream parameter also affects the likelihood contributions of subsequent trees through the conditional pseudo-observations generated recursively by the $h$-functions. Marginal and early-tree parameters can therefore receive additional information from several downstream components under the complete likelihood. The number of such downstream contributions decreases with tree depth and is zero for the final tree.

\begin{figure}[t]
    \centering
    \includegraphics[width=\textwidth]{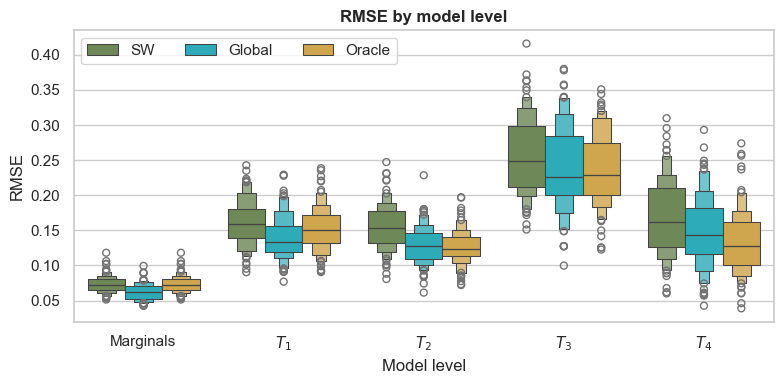}
    \caption{Root mean squared error of the posterior mean effects across the $100$ simulation replications for the three estimation strategies.}
    \label{fig:simulation_rmse}
\end{figure}

Figure~\ref{fig:simulation_rmse} shows that the increased concentration of the Global approximation is accompanied by an improvement in point estimation for several model levels. In particular, Global generally yields smaller RMSEs for the marginal and early-tree effects. The reduction in RMSE is, however, considerably smaller than the corresponding reduction in credible-interval width. Consequently, the increased concentration of the globally refined variational approximation is not matched by a proportional improvement in point-estimation accuracy, leading to the lower empirical coverage observed in Figure~\ref{fig:simulation_coverage}.

The stronger concentration of the Global approximation may also reflect a known limitation of variational inference. Variational posterior approximations can underestimate posterior dispersion and consequently produce overly narrow interval estimates, particularly when posterior dependencies are restricted by the variational family \citep{wang2005inadequacy}. The pronounced reduction in interval width under Global, relative to the more modest improvement in RMSE, is consistent with this phenomenon, although the present simulation does not isolate its contribution from the additional information introduced by joint optimization.

The Oracle comparison addresses a different aspect of the approximation. SW and Oracle have nearly identical interval widths, whereas the benefits of the oracle inputs appear primarily through reductions in estimation error at the deeper vine levels. Thus, recursively replacing the true conditional inputs by fitted pseudo-observations has comparatively little effect on the componentwise posterior dispersion, although the accumulated input error can affect point estimation in the deepest trees.

Taken together, the results show that the proposed tree-wise approximation remains close to the oracle benchmark while avoiding the computational burden of joint optimization. A final global refinement makes more complete use of the dependence of downstream likelihood contributions on upstream parameters and can improve point estimation, but under the factorized variational approximation in \eqref{eq:joint_variational_factorization} this additional information is accompanied by substantially more concentrated posterior approximations and poorer frequentist coverage for the upstream components.

\section{Application to meteorological data from the Netherlands}\label{sec:application}

We apply the proposed framework to a six-dimensional meteorological response observed at weather stations across the Netherlands. The data are obtained from the daily climatological station records of the Royal Netherlands Meteorological Institute (KNMI) \citep{knmi_daily} and cover the year 2025. We retain only stations with complete daily records over the observation period. The resulting data set contains 25 weather stations and $n=9125$ station-day observations. We treat the station-day response vectors as conditionally independent given the spatial and temporal covariates; residual serial dependence within stations and residual spatial dependence between stations are not modelled.

The response vector consists of daily mean wind speed ($Y_1$), daily mean temperature ($Y_2$), daily precipitation amount ($Y_3$), daily mean relative humidity ($Y_4$), sunshine percentage ($Y_5$), and global radiation ($Y_6$). In the original KNMI data, daily mean wind speed is reported in $0.1\,\mathrm{m/s}$, temperature in $0.1^\circ\mathrm{C}$, precipitation in $0.1\,\mathrm{mm}$, relative humidity and sunshine as percentages, and global radiation in $\mathrm{J/cm^2}$. Before estimation, wind speed is log-transformed, temperature is converted to degrees Celsius, precipitation trace values coded as $-1$ are set to zero and the resulting values are transformed as $\log(y+1)$, global radiation is transformed as $\log(y+1)$, and relative humidity and sunshine percentage are rescaled to the unit interval. The latter two variables are clipped to $[10^{-6},1-10^{-6}]$ to ensure that all observations lie in the interior of the support of the Beta distribution. All observed percentage values are strictly below $100\%$, so the upper truncation serves only as a numerical safeguard.

Wind speed, temperature, precipitation, and global radiation are modelled using Normal distributions with location and scale parameters, whereas relative humidity and sunshine percentage are modelled using Beta distributions with mean and total concentration parameters. The Normal margins are therefore applied to the transformed rather than the original positive-valued wind-speed, precipitation, and radiation measurements. In particular, the logarithmic transformations reduce their strong right skewness and place them on a scale more suitable for Gaussian modelling. For precipitation, this provides a parsimonious continuous working model for the transformed daily amount rather than a separate model for the occurrence and positive amount of precipitation. Normal location parameters use the identity link, positive scale and concentration parameters use the softplus transformation, and Beta means are mapped to the unit interval through the logistic transformation.

\subsection{Model specification}\label{sec:application_model}

We use latitude, longitude, and calendar time as covariates. Calendar time is rescaled to $[0,1]$ over the year and represented through a cyclic smooth. For every marginal and pair-copula parameter, the structured additive predictor is
\begin{align*}
\eta_{i,r,p}
=
\beta_{r,p,0}
+
f_{r,p}^{\mathrm{lat}}(\mathrm{lat}_i)
+
f_{r,p}^{\mathrm{lon}}(\mathrm{lon}_i)
+
f_{r,p}^{\mathrm{time}}(t_i)
+
f_{r,p}^{\mathrm{lat,lon}}(\mathrm{lat}_i,\mathrm{lon}_i),
\end{align*}
where $t_i\in[0,1]$ denotes normalized calendar time. The marginal latitude and longitude effects are represented by spline bases of dimension $6$, while their joint spatial contribution is completed by an anisotropic tensor-product interaction constructed from the corresponding $6\times6$ marginal bases. The cyclic temporal effect uses a basis of dimension $10$. The identifiability constraints and penalty reparameterization are those described in Appendix~\ref{sec:spline_constraints}. We use the same smoothing-variance prior and hyperparameter specification as in the simulation study.

For the application, we use AIC within the information-criterion-based selection framework introduced in Section~\ref{sec:vine_selection} to select both the vine structure and the pair-copula families. For every admissible edge, the candidate set consists of Independence, Gaussian, BB1, Frank, Gumbel, and Clayton copulas. Thus, both pair-copula family selection and the maximum spanning-tree construction are based on the AIC of the fitted covariate-dependent pair-copula regressions. The Gaussian dependence parameter is mapped to $(-1,1)$ using the hyperbolic tangent. Parameters constrained to be positive are obtained through the softplus transformation, whereas parameters constrained to exceed one use $1+\operatorname{softplus}(\cdot)$. The Frank parameter is left on its unconstrained scale. Thus, for the BB1 family, $\theta>0$ is obtained through the softplus transformation and $\delta>1$ through $1+\operatorname{softplus}(\cdot)$. All parameters of the selected pair copulas are assigned the same spatial and temporal predictor structure as the marginal parameters. In particular, the dependence structure is allowed to vary nonlinearly both over the Netherlands and throughout the year. All models in the application are implemented and estimated using the probabilistic programming framework \texttt{liesel} \citep{riebl2026lieselpythonframeworkgraphbased}, using the development branch \texttt{optima-ma-loss}. Further computational details are provided in Appendix~\ref{app:application_settings}.

\subsection{Selected vine structure}\label{sec:application_vine}

The selected regular vine is
\begin{align*}
T_1:\quad&
(1,3)_{\mathrm{BB1}},
\quad
(2,4)_{\mathrm{Gaussian}},
\quad
(3,6)_{\mathrm{Frank}},
\quad
(4,6)_{\mathrm{Gaussian}},
\quad
(5,6)_{\mathrm{BB1}},
\\
T_2:\quad&
(1,6;3)_{\mathrm{BB1}},
\quad
(2,6;4)_{\mathrm{Gaussian}},
\quad
(3,4;6)_{\mathrm{Gaussian}},
\quad
(4,5;6)_{\mathrm{Gaussian}},
\\
T_3:\quad&
(1,4;3,6)_{\mathrm{Gaussian}},
\quad
(2,3;4,6)_{\mathrm{Frank}},
\quad
(3,5;4,6)_{\mathrm{BB1}},
\\
T_4:\quad&
(1,2;3,4,6)_{\mathrm{Gaussian}},
\quad
(1,5;3,4,6)_{\mathrm{Frank}},
\\
T_5:\quad&
(2,5;1,3,4,6)_{\mathrm{Gumbel}}.
\end{align*}

\begin{figure}[H]
    \centering
    \includegraphics[width=0.95\textwidth]{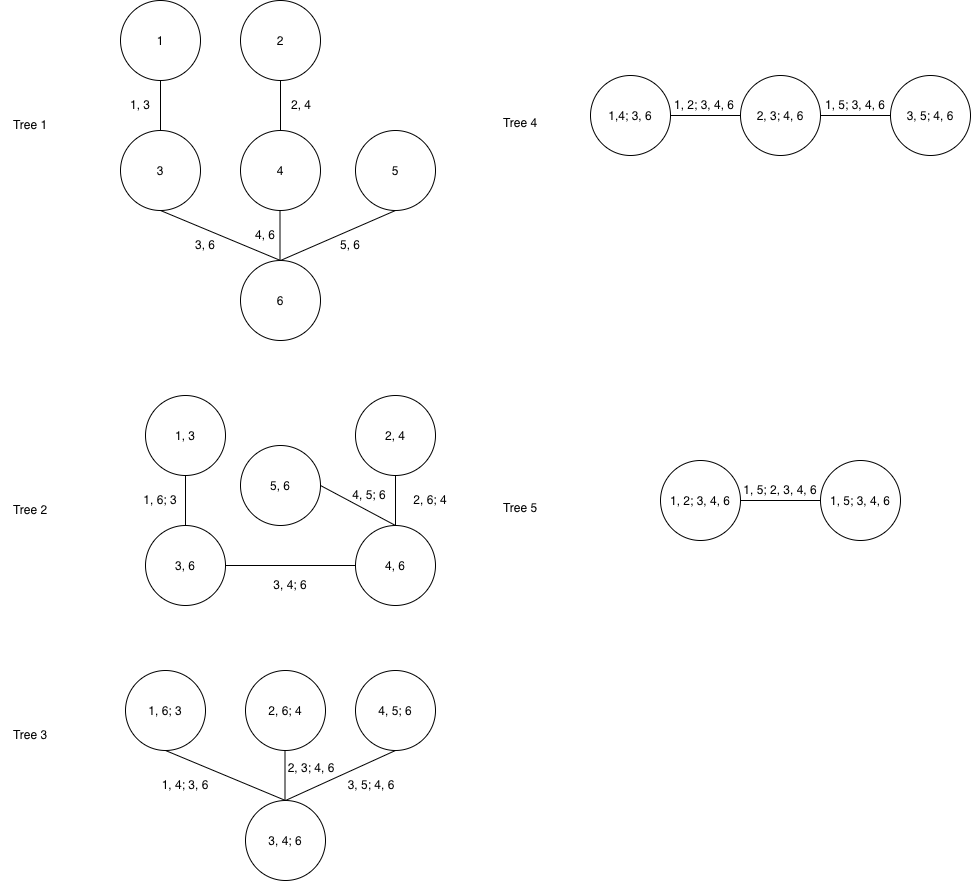}
    \caption{Graphical representation of the regular vine selected in the meteorological application. The response indices correspond to $1$: daily mean wind speed, $2$: daily mean temperature, $3$: daily precipitation amount, $4$: daily mean relative humidity, $5$: sunshine percentage, and $6$: global radiation. Nodes in $T_1$ correspond to response components, while nodes in subsequent trees correspond to edges of the preceding tree and are labelled by the associated conditioned variables and conditioning set.}
    \label{fig:application_vine}
\end{figure}

The selected dependence structure is displayed in Figure~\ref{fig:application_vine}.

The first tree admits a relatively direct meteorological interpretation. Global radiation $Y_6$ forms a central node and is connected to precipitation, relative humidity, and sunshine percentage. These variables are naturally linked through cloud cover and the surface radiation balance: radiation is closely related to sunshine duration, while cloudy and precipitating conditions are also associated with humidity and reduced incoming solar radiation. The remaining first-tree edges connect wind speed with precipitation and temperature with relative humidity, two further relationships that are consistent with the joint evolution of atmospheric conditions. Since the first tree consists of pairs that are unconditional with respect to the remaining response components, these edges represent the first-level dependence relationships favored by the AIC-based selection procedure.

The interpretation of subsequent trees is necessarily more conditional. In $T_2$, the model captures dependence between wind speed and global radiation conditional on precipitation, between temperature and global radiation conditional on relative humidity, between precipitation and relative humidity conditional on global radiation, and between relative humidity and sunshine conditional on global radiation. Higher trees continue to represent residual associations after successively conditioning on variables selected at preceding levels. In $T_4$, for example, wind speed and temperature are associated conditional on precipitation, relative humidity, and global radiation, while wind speed and sunshine percentage are related conditional on the same variables. The final tree represents the remaining dependence between temperature and sunshine percentage conditional on wind speed, precipitation, relative humidity, and global radiation. These higher-tree edges should therefore not be interpreted as isolated pairwise meteorological relationships, but rather as dependence remaining after the lower-tree structure has already been accounted for.

The selected pair-copula families also illustrate the motivation for using a vine rather than a single multivariate Gaussian copula. Of the $15$ pair copulas, seven are Gaussian, while the remaining eight consist of four BB1, three Frank, and one Gumbel copula. Non-Gaussian dependence is already prominent in the first tree, where the wind--precipitation and sunshine--radiation relationships are represented by BB1 copulas and the precipitation--radiation relationship by a Frank copula. It also persists throughout the higher trees. In $T_2$, the wind--radiation relationship conditional on precipitation is represented by a BB1 copula. In $T_3$, the temperature--precipitation relationship conditional on relative humidity and global radiation is represented by a Frank copula, while the precipitation--sunshine relationship under the same conditioning set is represented by a BB1 copula. A Frank copula is selected for the wind--sunshine relationship in $T_4$, and the final conditional temperature--sunshine relationship is represented by a Gumbel copula. In contrast, the truly multivariate distributional regression model of \citet{kock2023truly} represents the complete dependence structure through a Gaussian copula with a covariate-dependent correlation matrix. The selected vine indicates that a more flexible family specification is relevant in the present application, since different Gaussian and non-Gaussian copulas are favored for different unconditional and conditional pairs. The vine construction provides this flexibility while retaining a coherent six-dimensional joint distribution.

\subsection{Estimated effects}\label{sec:application_effects}

\begin{figure}[H]
    \centering
    \includegraphics[width=\textwidth]{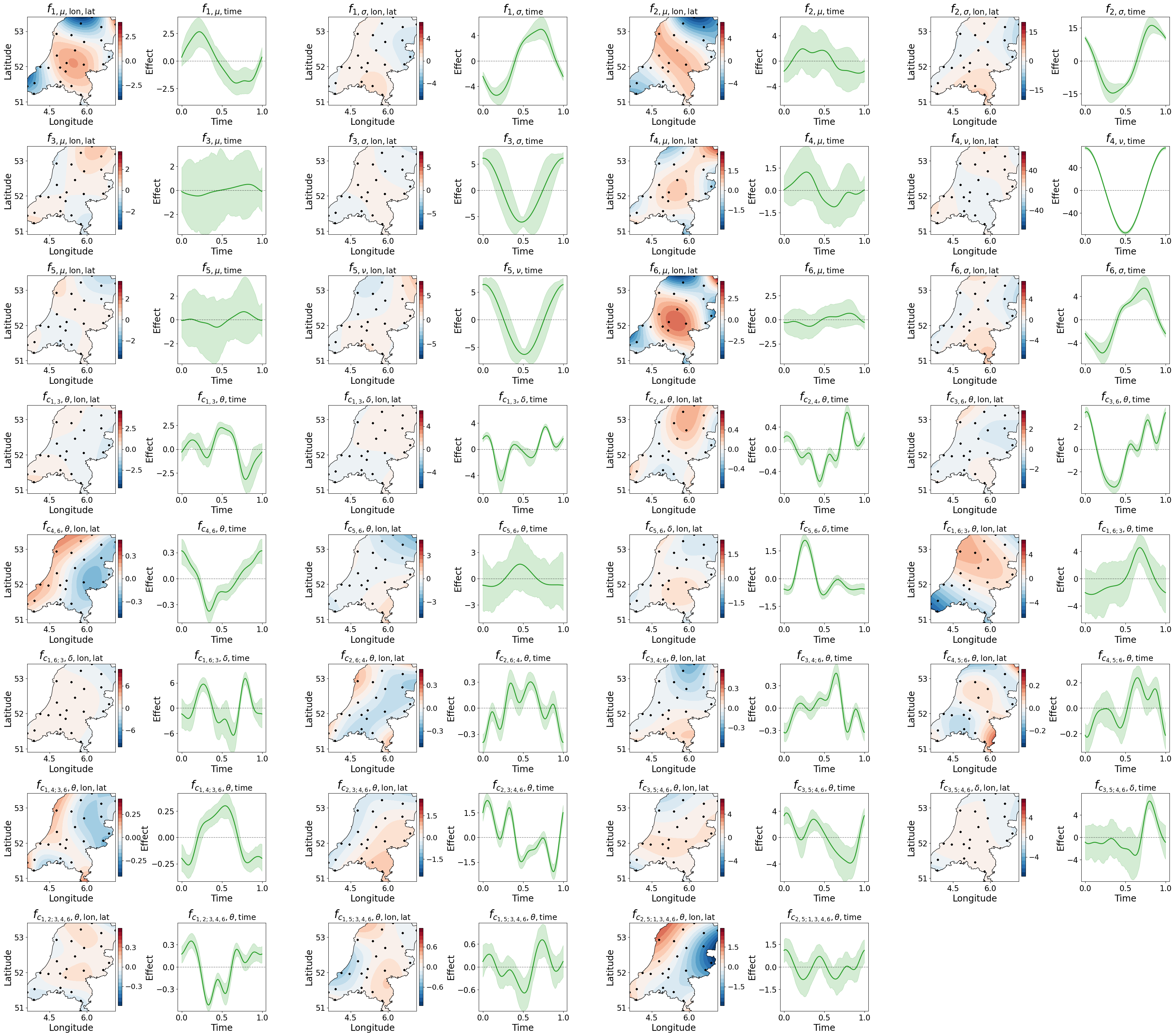}
    \caption{Estimated spatial and temporal effects in the meteorological application. For each marginal and pair-copula parameter, the spatial panel shows the posterior mean of the combined latitude, longitude, and anisotropic tensor-product interaction, with weather-station locations indicated by black points. Temporal panels show the posterior mean cyclic effect and pointwise $95\%$ credible intervals.}
    \label{fig:application_effects}
\end{figure}

Figure~\ref{fig:application_effects} summarizes the estimated spatial and temporal effects for every marginal and pair-copula parameter. For each parameter, the spatial panel displays the posterior mean of the combined spatial contribution
\begin{align*}
f_{r,p}^{\mathrm{space}}(\mathrm{lat},\mathrm{lon})
=
f_{r,p}^{\mathrm{lat}}(\mathrm{lat})
+
f_{r,p}^{\mathrm{lon}}(\mathrm{lon})
+
f_{r,p}^{\mathrm{lat,lon}}(\mathrm{lat},\mathrm{lon}),
\end{align*}
where $f_{r,p}^{\mathrm{lat,lon}}$ denotes the anisotropic tensor-product interaction. The surfaces are displayed only within the geographical boundary of the Netherlands, and the black points indicate the locations of the retained weather stations. The temporal panels show the posterior mean cyclic effect together with pointwise $95\%$ credible intervals. All effects are displayed on their corresponding additive-predictor scales.

The fitted marginal regressions exhibit clear nonlinear variation over both space and calendar time. The spatial surfaces show smooth but nonuniform geographical patterns across the Netherlands, while the temporal effects range from broad seasonal changes to more localized nonlinear variation throughout the year. Although the magnitude of the estimated effects differs considerably across parameters, the overall pattern supports the use of flexible spatial and cyclic smooths rather than restricting the meteorological covariates to linear effects.

More importantly for the multivariate model, substantial nonlinear covariate effects are also present in the dependence structure. Already among the five first-tree pair copulas, the estimated dependence parameters vary over both space and time, including the parameters of the selected BB1 and Frank copulas. The two parameters of the BB1 components can themselves display distinct covariate patterns, illustrating that covariates may alter different aspects of the same bivariate dependence structure. Clear spatial gradients and nonlinear temporal effects remain visible for several conditional pair copulas in the higher trees as well, including the BB1, Frank, and Gumbel components selected beyond the first tree.

The application therefore supports modelling covariate effects directly in the pair-copula parameters rather than treating the dependence structure as fixed after accounting for covariate-dependent margins. Flexible marginal regressions alone would not capture the remaining spatial and temporal variation visible in the fitted copulas. Conversely, the presence of both strongly structured and comparatively weak effects shows that the model is not merely reproducing a common spatial or seasonal pattern across all components: each marginal and pair-copula parameter is allowed to adapt separately to the data.

Taken together, the selected vine and the estimated effects illustrate the two main forms of flexibility provided by the proposed model. Dependence can differ qualitatively across pairs through the selection of different copula families, and the parameters within each selected family can vary nonlinearly with the covariates. Both features are relevant in this application: BB1, Frank, and Gumbel copulas are selected at multiple vine levels alongside Gaussian pair copulas, while spatial and temporal effects remain visible throughout the marginal and dependence components of the model.

\section{Conclusion}\label{sec:conclusion}

We have introduced a framework for multivariate structured additive distributional regression based on regular vine copulas. The construction provides dependence flexibility at two levels. Different vine edges may employ different pair-copula families, allowing symmetric, asymmetric, tail-dependent, and tail-independent forms of association to coexist within the same multivariate model. At the same time, every parameter of every selected pair copula may vary with observed covariates through a structured additive predictor. Structured additive distributional regression is therefore applied both to the conditional marginal distributions and to the pair-copula components of the multivariate dependence structure.

The resulting model can contain a large number of regression coefficients and smoothing parameters, while higher-tree likelihood contributions depend recursively on conditional probability integral transforms generated by preceding components. We address this structure through tree-wise stochastic variational inference. Marginal distributional regressions are estimated first, followed by pair-copula regressions sequentially along the vine trees, with fitted $h$-functions providing the conditional pseudo-observations required at subsequent levels. This reduces the full multivariate estimation problem to a sequence of marginal and bivariate distributional regressions.

The simulation study indicates that the proposed sequential approximation recovers nonlinear effects well and remains close to the oracle benchmark based on the true recursive conditional inputs. Differences become more visible in deeper vine trees, where errors from preceding levels can accumulate. Joint refinement under the complete likelihood can improve point estimation for some marginal and early-tree components by propagating information through the recursively constructed conditional probability integral transforms. In the investigated setting, however, this additional information is accompanied by substantially narrower variational intervals and lower frequentist coverage. The tree-wise procedure therefore retains competitive point estimation while providing more reliable uncertainty quantification in this experiment.

When the vine is not specified in advance, we complement the estimation procedure with a sequential strategy for selecting both the structure and the pair-copula families. Each admissible edge is fitted under the candidate families and evaluated using an information criterion based on effective degrees of freedom. For each candidate edge, the family minimizing the chosen criterion is retained, and its negative information-criterion value is used as the edge weight in the maximum spanning-tree problem. The resulting procedure therefore bases both family and structure selection directly on the fitted covariate-dependent pair-copula regressions.

The meteorological application illustrates both sources of dependence flexibility in a six-dimensional setting. Using AIC within the proposed selection framework, the selected vine combines Gaussian, BB1, Frank, and Gumbel pair copulas. At the same time, the estimated pair-copula parameters exhibit pronounced nonlinear spatial and temporal variation, both in the first tree and for several conditional pair copulas at higher levels. The fitted model therefore allows dependence to vary both in its pair-specific copula family and through covariate-dependent parameters within each selected family.

An important direction for future work concerns uncertainty propagation across the sequential vine construction. The present tree-wise procedure treats fitted quantities from preceding stages as fixed when estimating subsequent components. Developing scalable variational approximations that retain more of the dependence between these components, while avoiding the strong posterior concentration observed under the current global refinement, could further improve uncertainty quantification without sacrificing the computational advantages of sequential estimation.

Overall, regular vines and structured additive distributional regression provide complementary forms of flexibility: the vine determines how multivariate dependence is decomposed and which pair-copula family is used for each edge, while structured additive predictors determine how the parameters of these pairwise dependence models vary with covariates. The proposed stochastic variational inference framework makes this combination computationally feasible for multivariate distributional regression models with many marginal and pair-copula components.

\bibliographystyle{agsm}
\bibliography{bibliography}

@article{kneib2021,
  author = {Kneib, Thomas and Silbersdorff, Alexander and S{\"a}fken, Benjamin},
  title = {Rage Against the Mean---A Review of Distributional Regression Approaches},
  journal = {Econometrics and Statistics},
  year = {2023},
  volume = {26},
  pages = {99--123},
  doi = {10.1016/j.ecosta.2021.07.006}
}

@article{kock2023truly,
  author = {Kock, Lucas and Klein, Nadja},
  title = {Truly Multivariate Structured Additive Distributional Regression},
  journal = {Journal of Computational and Graphical Statistics},
  year = {2025},
  volume = {34},
  number = {4},
  pages = {1189--1201},
  doi = {10.1080/10618600.2024.2434181}
}

@article{marra2017bivariate,
  title={Bivariate copula additive models for location, scale and shape},
  author={Marra, Giampiero and Radice, Rosalba},
  journal={Computational Statistics \& Data Analysis},
  volume={112},
  pages={99--113},
  year={2017},
  publisher={Elsevier}
}

@article{bedford2001probability,
  title={Probability density decomposition for conditionally dependent random variables modeled by vines},
  author={Bedford, Tim and Cooke, Roger M},
  journal={Annals of Mathematics and Artificial intelligence},
  volume={32},
  number={1},
  pages={245--268},
  year={2001},
  publisher={Springer}
}

@article{bedford2002vines,
  title={Vines--a new graphical model for dependent random variables},
  author={Bedford, Tim and Cooke, Roger M},
  journal={The Annals of statistics},
  volume={30},
  number={4},
  pages={1031--1068},
  year={2002},
  publisher={Institute of Mathematical Statistics}
}

@article{aas2009,
  author = {Aas, Kjersti and Czado, Claudia and Frigessi, Arnoldo and Bakken, Henrik},
  title = {Pair-Copula Constructions of Multiple Dependence},
  journal = {Insurance: Mathematics and Economics},
  year = {2009},
  volume = {44},
  number = {2},
  pages = {182--198},
  doi = {10.1016/j.insmatheco.2007.02.001}
}

@book{joe2014dependence,
  title={Dependence modeling with copulas},
  author={Joe, Harry},
  year={2014},
  publisher={CRC press}
}

@article{czado2022vine,
  title={Vine copula based modeling},
  author={Czado, Claudia and Nagler, Thomas},
  journal={Annual Review of Statistics and Its Application},
  volume={9},
  pages={453--477},
  year={2022},
  publisher={Annual Reviews}
}

@article{griesbauer2026stepwise,
  author = {Griesbauer, Elisabeth and R{\o}nneberg, Leiv and Frigessi, Arnoldo and Czado, Claudia and Hob{\ae}k Haff, Ingrid},
  title = {Stepwise Variational Inference with Vine Copulas},
  journal = {arXiv preprint arXiv:2603.22959},
  year = {2026},
  doi = {10.48550/arXiv.2603.22959}
}

@article{dissmann2013selecting,
  title={Selecting and estimating regular vine copulae and application to financial returns},
  author={Dissmann, Jeffrey and Brechmann, Eike C and Czado, Claudia and Kurowicka, Dorota},
  journal={Computational statistics \& data analysis},
  volume={59},
  pages={52--69},
  year={2013},
  publisher={Elsevier}
}

@inproceedings{sklar1959fonctions,
  title={Fonctions de r{\'e}partition {\`a} n dimensions et leurs marges},
  author={Sklar, M},
  booktitle={Annales de l'ISUP},
  volume={8},
  number={3},
  pages={229--231},
  year={1959}
}

@article{patton2006modelling,
  author = {Patton, Andrew J.},
  title = {Modelling Asymmetric Exchange Rate Dependence},
  journal = {International Economic Review},
  year = {2006},
  volume = {47},
  number = {2},
  pages = {527--556},
  doi = {10.1111/j.1468-2354.2006.00387.x}
}

@article{stoeber2013,
  author = {St{\"o}ber, Jakob and Joe, Harry and Czado, Claudia},
  title = {Simplified Pair Copula Constructions---Limitations and Extensions},
  journal = {Journal of Multivariate Analysis},
  year = {2013},
  volume = {119},
  pages = {101--118},
  doi = {10.1016/j.jmva.2013.04.014}
}

@article{nagler2025simplified,
  title={Simplified vine copula models: state of science and affairs},
  author={Nagler, Thomas},
  journal={Risk Sciences},
  volume={1},
  pages={100022},
  year={2025},
  publisher={Elsevier}
}

@article{vatter2018generalized,
  title={Generalized additive models for pair-copula constructions},
  author={Vatter, Thibault and Nagler, Thomas},
  journal={Journal of Computational and Graphical Statistics},
  volume={27},
  number={4},
  pages={715--727},
  year={2018},
  publisher={Taylor \& Francis}
}

@article{rigby2005generalized,
  title={Generalized additive models for location, scale and shape},
  author={Rigby, Robert A and Stasinopoulos, D Mikis},
  journal={J Royal Stat Soc C},
  volume={54},
  number={3},
  pages={507--554},
  year={2005},
  publisher={Wiley Online Library}
}

@article{hans2023boosting,
  title={Boosting distributional copula regression},
  author={Hans, Nicolai and Klein, Nadja and Faschingbauer, Florian and Schneider, Michael and Mayr, Andreas},
  journal={Biometrics},
  volume={79},
  number={3},
  pages={2298--2310},
  year={2023},
  publisher={Oxford University Press}
}

@article{klein2015bayesian,
  title={Bayesian structured additive distributional regression for multivariate responses},
  author={Klein, Nadja and Kneib, Thomas and Klasen, Stephan and Lang, Stefan},
  journal={Journal of the Royal Statistical Society Series C: Applied Statistics},
  volume={64},
  number={4},
  pages={569--591},
  year={2015},
  publisher={Oxford University Press}
}

@article{klein2016simultaneous,
  title={Simultaneous inference in structured additive conditional copula regression models: a unifying Bayesian approach},
  author={Klein, Nadja and Kneib, Thomas},
  journal={Statistics and Computing},
  volume={26},
  number={4},
  pages={841--860},
  year={2016},
  publisher={Springer}
}

@article{klein2022multivariate,
  title={Multivariate conditional transformation models},
  author={Klein, Nadja and Hothorn, Torsten and Barbanti, Luisa and Kneib, Thomas},
  journal={Scandinavian Journal of Statistics},
  volume={49},
  number={1},
  pages={116--142},
  year={2022},
  publisher={Wiley Online Library}
}

@article{muschinski2024cholesky,
  title={Cholesky-based multivariate Gaussian regression},
  author={Muschinski, Thomas and Mayr, Georg J and Simon, Thorsten and Umlauf, Nikolaus and Zeileis, Achim},
  journal={Econometrics and Statistics},
  volume={29},
  pages={261--281},
  year={2024},
  publisher={Elsevier}
}

@article{blei2017variational,
  title={Variational inference: A review for statisticians},
  author={Blei, David M and Kucukelbir, Alp and McAuliffe, Jon D},
  journal={J Am Stat Assoc},
  volume={112},
  number={518},
  pages={859--877},
  year={2017},
  publisher={Taylor \& Francis}
}

@article{hoffman2013stochastic,
  title={Stochastic variational inference},
  author={Hoffman, Matthew D and Blei, David M and Wang, Chong and Paisley, John},
  journal={J Machine Learning Res},
  volume={14},
  number={1},
  pages={1303--1347},
  year={2013},
  publisher={JMLR. org}
}

@article{callegher2024stochastic,
  title={Stochastic Variational Inference for Structured Additive Distributional Regression},
  author={Callegher, Gianmarco and Kneib, Thomas and S{\"o}ding, Johannes and Wiemann, Paul},
  journal={arXiv preprint arXiv:2412.10038},
  year={2024}
}

@inproceedings{wang2005inadequacy,
  title={Inadequacy of interval estimates corresponding to variational Bayesian approximations},
  author={Wang, Bo and Titterington, Donald M},
  booktitle={International workshop on artificial intelligence and statistics},
  pages={373--380},
  year={2005},
  organization={PMLR}
}

@misc{knmi_daily,
  author = {{Royal Netherlands Meteorological Institute (KNMI)}},
  title = {Daggegevens van weerstations},
  howpublished = {\url{https://daggegevens.knmi.nl/klimatologie/daggegevens}},
  note = {Accessed 17 August 2026}
}

@misc{riebl2026lieselpythonframeworkgraphbased,
      title={Liesel: A Python Framework for Graph-Based Bayesian Modeling and Customizable MCMC with Support for Generalized Additive Models}, 
      author={Hannes Riebl and Johannes Brachem and Thomas Kneib and Gianmarco Callegher and Paul F. V. Wiemann},
      year={2026},
      eprint={2209.10975},
      archivePrefix={arXiv},
      primaryClass={stat.CO},
      url={https://arxiv.org/abs/2209.10975}, 
}

@article{wood2016smoothing,
  title   = {Smoothing Parameter and Model Selection for General Smooth Models},
  author  = {Wood, Simon N. and Pya, Natalya and S{\"a}fken, Benjamin},
  journal = {Journal of the American Statistical Association},
  volume  = {111},
  number  = {516},
  pages   = {1548--1563},
  year    = {2016},
  doi     = {10.1080/01621459.2016.1180986}
}

\appendix
\section{Appendix}

\subsection{Spline constraints and penalty reparameterization}\label{sec:spline_constraints}

This section records the implementation details used to make the additive predictors identifiable and to express each spline effect in the penalized/unpenalized form used by the Gaussian priors. These steps are straightforward but included here for completeness.

For a smooth effect $f_{i,d,p,j}=\X_{d,p,j}\bs\beta_{d,p,j}$, identifiability is enforced by centering the fitted effect over the observed covariates,
\begin{align*}
\sum_{i=1}^n f_{i,d,p,j}^{\rm new}(x_{i,d,p,j},\bs\beta_{d,p,j})=0.
\end{align*}
Let $\C_{d,p,j}$ be the vector of column means of $\X_{d,p,j}$. A QR decomposition of $\C_{d,p,j}$ gives an orthogonal basis $\Q_{d,p,j}=[\mathbf Z^{(a)}_{d,p,j},\mathbf Z^{(b)}_{d,p,j}]$, where $\mathbf Z^{(b)}_{d,p,j}$ spans the subspace satisfying the centering constraint. We therefore use the constrained design and penalty
\begin{align*}
\X_{d,p,j}^{\rm new}=\X_{d,p,j}\mathbf Z^{(b)}_{d,p,j}, \qquad
\K_{d,p,j}^{\rm new}=\left(\mathbf Z^{(b)}_{d,p,j}\right)^\T\K_{d,p,j}\mathbf Z^{(b)}_{d,p,j}.
\end{align*}
For one distributional parameter $p$ in component $d$, the full design and penalty matrices are then
\begin{align*}
\X_{d,p}=\left[\mathbf 1,\X_{d,p,1}^{\rm new},\dots,\X_{d,p,J_{d,p}}^{\rm new}\right].
\end{align*}
\begin{align*}
\bs K_{d,p}(\bs\tau_{d,p}^2)
=
\operatorname{blockdiag}
\left\{
0,
\frac{1}{\tau_{d,p,1}^2}\bs K^{\mathrm{new}}_{d,p,1},
\dots,
\frac{1}{\tau_{d,p,J_{d,p}}^2}\bs K^{\mathrm{new}}_{d,p,J_{d,p}}
\right\}.
\end{align*}
To separate penalized and unpenalized components, let
\begin{align*}
\K^{\rm new}_{d,p,j}
=
\U_{d,p,j}
\begin{pmatrix}
\boldsymbol\Lambda_{d,p,j}^{+} & 0\\
0 & 0
\end{pmatrix}
\U_{d,p,j}^\T,
\end{align*}
where $\boldsymbol\Lambda_{d,p,j}^{+}=\operatorname{diag}(\lambda_1,\dots,\lambda_r)$ contains the positive eigenvalues of $\K^{\rm new}_{d,p,j}$. Writing $\U_{d,p,j}=[\U^{+}_{d,p,j},\U^{0}_{d,p,j}]$, with $\U^{0}_{d,p,j}$ spanning the null space of the penalty, we define
\begin{align*}
\X_{d,p,j}^{\rm penalized}
&=
\X_{d,p,j}^{\rm new}\U^{+}_{d,p,j}
\left(\boldsymbol\Lambda_{d,p,j}^{+}\right)^{-1/2},\\
\X_{d,p,j}^{\rm unpenalized}
&=
\X_{d,p,j}^{\rm new}\U^{0}_{d,p,j}.
\end{align*}
Thus
\begin{align*}
\X_{d,p,j}^{\rm transformed}
=
\left[
\X_{d,p,j}^{\rm penalized},
\X_{d,p,j}^{\rm unpenalized}
\right].
\end{align*}
Under this reparameterization, the penalty becomes the identity on the penalized block and zero on the unpenalized block, giving the usual mixed-model representation of spline effects used in the structured additive predictor.

\subsection{Estimation algorithm}\label{sec:estimation_algorithm}

The complete procedure is summarized in Algorithm~\ref{alg:sequential_vine_svi}. For every admissible candidate edge, all candidate pair-copula families are fitted once and evaluated using the chosen information criterion. The family minimizing this criterion determines the edge weight used in the maximum spanning-tree problem. After the tree is selected, the corresponding already fitted pair-copula regressions are retained and used to construct the conditional pseudo-observations for the next tree.

\begin{algorithm}[t]
\caption{Sequential stochastic variational inference for vine copula distributional regression}
\label{alg:sequential_vine_svi}
\begin{algorithmic}[1]
\Require Data $\{(\y_i,\x_i)\}_{i=1}^n$, candidate pair-copula families $\mathfrak C$, and information criterion $\operatorname{IC}$
\Ensure Marginal fits, selected regular vine $\widehat{\mathcal V}$, selected pair-copula families, pair-copula fits, and conditional pseudo-observations

\State Initialize the conditional-uniform store $\mathcal S\leftarrow\emptyset$

\For{$d=1,\dots,D$}
    \State Estimate marginal distributional regression $d$ by maximizing the componentwise ELBO in \eqref{eq:component_elbo}
    \State Compute $\widehat u_{i,d}$ according to \eqref{eq:fitted_marginal_pits}, for $i=1,\dots,n$
    \State Store $\{\widehat u_{i,d}\}_{i=1}^n$ under the key $(d,\emptyset)$ in $\mathcal S$
\EndFor

\For{$m=1,\dots,D-1$}
    \If{$m=1$}
        \State Set the graph nodes to $\{1,\dots,D\}$
        \State Construct all unconditional candidate edges in $\mathcal C_m$
    \Else
        \State Set the graph nodes to the selected edges of $\widehat T_{m-1}$
        \State Construct $\mathcal C_m$ from all candidate edges satisfying the proximity condition in Definition~\ref{def:regular_vine}
    \EndIf

    \For{every candidate edge $e_m\in\mathcal C_m$}
        \State Retrieve $\widehat u_{i,j_{e_m}\mid D_{e_m}}$ and $\widehat u_{i,k_{e_m}\mid D_{e_m}}$ from $\mathcal S$, for $i=1,\dots,n$

        \For{every candidate family $c\in\mathfrak C$}
            \State Fit the pair-copula distributional regression for edge $e_m$ under family $c$
            \State Compute $\widehat{\ell}_{e_m,c}$ according to \eqref{eq:candidate_loglik}
            \State Compute $\widehat{\operatorname{df}}_{e_m,c}$ from the fitted variational covariance and penalty matrix as described in Section~\ref{sec:vine_selection}
            \State Compute $\operatorname{IC}_{e_m,c}$ using the chosen information criterion
        \EndFor

        \State Determine $\widehat c_{e_m}$ according to \eqref{eq:candidate_family_selection}
        \State Set $w_{e_m}\leftarrow-\operatorname{IC}_{e_m,\widehat c_{e_m}}$ according to \eqref{eq:candidate_edge_weight}
    \EndFor

    \State Select $\widehat T_m$ by solving the maximum spanning-tree problem in \eqref{eq:vine_mst}

    \For{every selected edge $e_m\in E(\widehat T_m)$}
        \State Retain the fitted regression for edge $e_m$ under family $\widehat c_{e_m}$
        \State Compute the fitted conditional transforms in \eqref{eq:fitted_h_first}--\eqref{eq:fitted_h_second}
        \State Store the resulting conditional pseudo-observations in $\mathcal S$
    \EndFor
\EndFor

\State Set $\widehat{\mathcal V}=(\widehat T_1,\dots,\widehat T_{D-1})$
\State \Return $\widehat{\mathcal V}$, the marginal and pair-copula fits, the selected families, and $\mathcal S$
\end{algorithmic}
\end{algorithm}

\subsection{Simulation settings}\label{app:simulation_settings}

This section provides the computational details for the simulation study in Section~\ref{sec:simulation}.

Each nonlinear covariate effect is represented using a spline basis with $10$ regression coefficients, spline order $3$, and penalty order $r=2$. The centering constraint and penalty reparameterization are implemented as described in Section~\ref{sec:spline_constraints}. In the simulation study, all smoothing variances are assigned an $\operatorname{Inverse\text{-}Gamma}(1,1)$ hyperprior, with concentration and scale both equal to $1$.

All stochastic variational optimizations are run for at most $2000$ epochs using mini-batches of size $B=1000$. The SW and Oracle procedures use Adam with learning rate $0.01$. The final joint optimization of the Global estimator is initialized at the completed SW solution and uses Adam with the smaller learning rate $0.001$.

At every optimization step, the Monte Carlo approximation of the ELBO and its gradient is based on $32$ samples from the variational distribution. Convergence is monitored using an exponentially smoothed moving average of the ELBO with smoothing parameter $\rho=0.1$. Early stopping is used with a patience of $200$ epochs.

After estimation, posterior summaries of the nonlinear effects are computed from $100$ draws from the fitted variational distribution, as described in Section~\ref{sec:simulation_metrics}.

For the Global estimator, the complete variational distribution is initialized using the variational parameters obtained from the preceding SW fit. During the final joint optimization, the conditional pseudo-observations entering the higher vine trees are recomputed from the current parameter values whenever the complete likelihood is evaluated.

\subsection{Application settings}\label{app:application_settings}

This section provides the computational details for the meteorological application in Section~\ref{sec:application}. The latitude and longitude effects are constructed from spline bases with dimension $6$, and their tensor-product interaction uses the corresponding $6\times6$ construction. Calendar time is represented by a cyclic spline with basis dimension $10$. Centering and penalty reparameterization are performed as described in Section~\ref{sec:spline_constraints}. As in the simulation study, all smoothing variances are assigned an $\operatorname{Inverse\text{-}Gamma}(1,1)$ hyperprior.

All stochastic variational optimizations are run for at most $4000$ epochs using mini-batches of size $B=1000$. At every optimization step, the Monte Carlo approximation of the ELBO and its gradient is based on $8$ samples from the variational distribution. Optimization is performed with Adam using learning rate $0.001$, preceded by global-norm gradient clipping at $1$ and replacement of numerical NaN values by zero.

Convergence is monitored using an exponentially smoothed moving average of the ELBO with smoothing parameter $\rho=0.1$. Early stopping is used with a patience of $200$ epochs. Absolute and relative convergence tolerances are both set to zero, so termination before the maximum number of epochs is determined by the patience criterion.

For the vine-selection procedure, the candidate set consists of Independence, Gaussian, BB1, Frank, Gumbel, and Clayton pair-copula families. We use AIC within the information-criterion-based selection framework of Section~\ref{sec:vine_selection}; both pair-copula family selection and the maximum spanning-tree construction are therefore based on the AIC computed using the effective degrees of freedom defined there. Once estimation is complete, posterior summaries and the effects displayed in Figure~\ref{fig:application_effects} are computed from $100$ draws from the fitted variational distributions.

\end{document}